\documentclass[aps,prd,nofootinbib,12pt]{revtex4}
\usepackage{epsfig}
\usepackage{graphicx}
\usepackage[font=small,skip=0pt,justification=raggedright]{caption}
\usepackage{amsmath,amssymb,mathrsfs}
\usepackage{verbatim}
\newcounter{fig}   \newcommand{\lbfig}[1]{\refstepcounter{fig}
\label{#1} }

\newcommand{\bea}{\begin{eqnarray}}
\newcommand{\eea}{\end{eqnarray}}
\newcommand{\be}{\begin{equation}}
\newcommand{\ee}{\end{equation}}

\newcommand{\re}[1]{(\ref{#1})}

\newcommand{\eqn}{\begin{eqnarray}}
\newcommand{\eqnx}{\end{eqnarray}}

\date{\today}

\begin{document}
\title{Q-ball stress stability criterion in the $U(1)$ gauged scalar theories}
\author{Victor~Loiko}
\affiliation{ Department of Theoretical Physics and Astrophysics,
Belarusian State University, Minsk 220004, Belarus}
\author{Ya.~Shnir}
\affiliation{BLTP, JINR, Dubna 141980, Moscow Region, Russia}
\affiliation{Institute of Physics, University of Oldenburg,
Oldenburg D-26111, Germany}

\begin{abstract}
We study the  energy-momentum tensor of the spherically symmetric $U(1)$ gauged Q-ball configurations
in the two-component Fridberg-Lee-Sirlin-Maxwell model, and in the one-component scalar model  with a sixtic potential.
We evaluate the distributions of the corresponding shear forces and pressure and study the
stability criteria for  these solutions. We present the results of numerical simulations in both models, explicitly
demonstrating that the electrostatic repulsion may destabilize
the $U(1)$ gauged Q-balls.
However, in the limiting case of the Fridberg-Lee-Sirlin-Maxwell model with a long ranged real scalar component,
the gauged Q-balls always remain stable.

\end{abstract}

\maketitle

\emph{Dedicated to the memory of Maxim Polyakov
}
\section{Introduction}
Q-balls are non-topological solitons, spacially localized
field configurations with finite energy in the flat 3+1
dimensional Minkovski spacetime. They represent time-dependent lumps of a
complex scalar field with a stationary oscillating phase
\cite{Rosen,Friedberg:1976me,Coleman:1985ki}. Such solutions
may exist in models
possessing an unbroken continuous
global symmetry,
typical examples are the model with a single complex scalar field and a suitable
non-renormalizable self-interaction potential \cite{Coleman:1985ki}, and so-called Friedberg-Lee-Sirlin two-component model with a symmetry breaking potential \cite{Friedberg:1976me}.
The Q-balls carry a Noether charge $Q$ associated with the $U(1)$ symmetry
(for a review, see, e.g. \cite{Lee:1991ax,Shnir2018,Radu:2008pp}), they can be considered as condensates of a large number of the field quanta which correspond to an
extremum of the energy effective energy functional for a fixed value of the charge $Q$.  The charge $Q$
also can be interpreted as the particle number. Certainly,
there is a similarity between the Q-balls and their non-relativistic
analogues, non-topological lumps in the Bose-Einstein condensate \cite{Enqvist:2003zb}.

Q-balls have received a lot of attention during the last decades, it was
suggested that such configuration may be formed in a primordial phase transition contributing to various scenario of
the evolution of the early Universe \cite{Frieman:1989bx,Kusenko:1997hj},
in particular, acting as a possible catalyst
for baryogenesis \cite{Affleck:1984fy,Enqvist:1997si}. The Q-balls also are considered
as candidates for dark matter \cite{Kusenko:1997si}, they may occur in the minimal
supersymmetric generalization of the Standard Model with the global charge $Q$ being identified with baryon or lepton number
\cite{Kusenko:1997zq,Campanelli:2009su,Campanelli:2007um}.

The local $U(1)$ symmetry of a model supporting Q-balls can be
promoted to be a local gauge symmetry, it corresponds to the
gauged Q-balls \cite{Lee:1988ag,Anagnostopoulos:2001dh,Gulamov:2013cra,Gulamov:2015fya,Nugaev:2019vru,Loiko:2019gwk}.
Notably, Q-ball configurations in the $U(1)$-gauged model of complex scalar field
with minimal electromagnetic coupling was considered already in the second of the pioneering papers by
Rosen \cite{Rosen}.

There are some important differences between the gauged and ungauged Q-ball solutions.
Spherically symmetric Q-balls with global $U(1)$ symmetry exist only in a certain angular frequency range,
$\omega\in [\omega_{min},\omega_{max}]$, determined by the properties of the potential. Both mass and charge of the
ungauged Q-balls diverge at both limiting values of the frequency,  there are two branches of solutions, bifurcating at
the minimal charge and mass.  On the other hand, both the energy and the charge of  Q-balls with local $U(1)$ symmetry
remain finite for all allowed values of the angular frequency \cite{Gulamov:2013cra,Gulamov:2015fya,Nugaev:2019vru}. The $U(1)$ gauged Q-balls
form two branches of solutions, they merge at some minimal value of the angular frequency \cite{Lee:1988ag}.

It was pointed out that the presence of the Abelian gauge field may affect
the properties of the soliton, as the gauge coupling increases, the electromagnetic repulsion
between the scalar particles destabilizes the configuration.
The problem of stability of the Q-balls has been analyzed in a number of works, see e.g.,
\cite{Lee:1991ax,Lee:1991bn,PaccettiCorreia:2001wtt,Sakai:2007ft}. It was shown that
for some range of values of the parameters of the model there are stable, metastable and unstable solutions.
In particular, Q-balls may be unstable with respect to linear perturbations of the fields, or because of non-linear
effects.  Q-balls with global $U(1)$ symmetry are stable along the lower branch, as the mass  of the configuration remains smaller
than the mass of free scalar particles with charge $Q$.

Another interesting approach was proposed in \cite{Mai:2012yc,Mai:2012cx}, it was suggested to study the
matrix elements of energy-momentum tensor and related spatial distributions of the forces acting in the interior of the configuration.
This approach is inspired by the study of the
form factors of the energy-momentum tensor of hadrons \cite{Polyakov:2002yz,Perevalova:2016dln,Polyakov:2018zvc} and evaluation of the
corresponding D-term, the quantity
which is related to the spacial deformations of the system \cite{Polyakov:2002yz,Perevalova:2016dln}.
It was shown that all finite energy Q-ball solutions with local $U(1)$ symmetry
satisfy certain criteria for the distribution of the shear force and the
pressure \cite{Mai:2012yc}.

A main objective of this paper is to study the energy-momentum tensor of the gauged spherically symmetric Q-balls
in the one-component gauged scalar model with a sixtic potential and in the
two-component Friedberg-Lee-Sirlin-Maxwell model.
In what follows, we begin by summarizing the properties of the gauged
Q-balls in flat space leading to a discussion of the stress tensor and the problem of the
distribution of the shear forces and pressure acting on the $U(1)$ gauged configuration.
Numerical results are presented in Section II, where we
discuss the stability condition of the gauged Q-balls which follow from the conservation of the
energy momentum tensor of the systems. Conclusions and remarks are
formulated in Section III.

\section{$U(1)$ gauged Q-balls}
\subsection{$U(1)$ gauged Q-balls in the model with sixtic potential}
First, we consider $U(1)$ gauged (3+1)-dimensional
self-interacting complex scalar field $\phi$, minimally
interacting with the Abelian gauge field $A_\mu$
\cite{Lee:1988ag}. The model is described by the Lagrangian \be
L^{(I)}=  -\frac14 F_{\mu\nu}F^{\mu\nu}+ D_\mu\phi D^\mu\phi^* -
U(|\phi|) \, , \label{lag-Coleman} \ee where $D_\mu = \partial_\mu
+igA_\mu$ denotes the covariant derivative, $A_\mu$ is a
four-potential, $g$ is the gauge coupling constant, the
electromagnetic field strength tensor is $F_{\mu\nu}=\partial_\mu
A_\nu-\partial_\nu A_\mu$ and the $U(1)$ invariant
self-interacting potential of the complex scalar field is
\cite{Lee:1988ag,Volkov:2002aj,Kleihaus:2005me} \be
\label{potential} U(|\phi|)=a|\phi|^2 - b |\phi|^4 + |\phi|^6 \, ,
\ee with the usual choice of the positive parameters  $a=1.1$ and
$b=2$. In such a case non-topological soliton solutions of this
model  exist only in a certain frequency range $\omega_{min} \le
\omega \le \omega_{max}$ \cite{Volkov:2002aj,Kleihaus:2005me},
where the upper limit $ \omega_{max}=\sqrt{\frac12 U^{\prime
\prime}(0)} =a $ corresponds to the mass of scalar excitation. In
the non-gauged case, $g=0$, the lower bound is given by the
condition
$$
\omega_{min}^2 =\frac{U(\phi_0)}{\phi_0^2} = a-\frac{b^2}{4}
$$
where $\phi_0$ is a minimum of the potential \re{potential}.
In the gauged theory \re{lag-Coleman} the minimal allowed value of the angular frequency is increasing as the gauge coupling grows,
at some critical value of the coupling the electrostatic repulsion becomes too strong for Q-ball to exists \cite{Lee:1988ag,Lee:1991bn,Anagnostopoulos:2001dh,Gulamov:2013cra,Gulamov:2015fya,Nugaev:2019vru}.

The model \re{lag-Coleman} is invariant with respect to the local $U(1)$ gauge transformations
\be
\phi \to \tilde \phi = e^{i \alpha(x)  }\phi, \quad A_\mu \to \tilde A_\mu = A_\mu  - \partial_\mu \alpha(x) \, ,
\ee
the associated Noether current is
\be
j_{\mu} = i \left(\phi^* D_\mu \phi  - \phi D_\mu \phi^* \right) ,
\label{Nother-curr}
\ee
and the conserved charge is
\be
Q=i\int d^3x j_0 \, .
\label{Nother-charge}
\ee

The Euler-Lagrange equation of the model \re{lag-Coleman} for the scalar field is
\be
D_\mu D^\mu \phi = -a \phi +2 b |\phi|^2 \phi - 3 |\phi|^4 \phi =0\, ,
\label{colemal_fe}
\ee
it is supplemented by the Maxwell equation
\be
\partial_\mu F^{\mu\nu} = g j^\nu
\label{max}
\ee
with the current $j^\nu$ \re{Nother-curr} as a source.

Further, the symmetrized  energy momentum tensor of the model \re{lag-Coleman} reads
\be
T_{\mu\nu}=  -\eta_{\mu\nu} \left(D_\rho\phi D^\rho \phi^* +\frac14 F_{\rho\sigma} F^{\rho\sigma}
+ U(|\phi|)\right)
+ \left(D_\mu\phi D_\nu \phi^* + D_\mu\phi^* D_\nu \phi\right) + \eta^{\rho \sigma}
F_{\mu \rho} F_{\nu \sigma}
\label{T_Coleman}
\ee
where $\eta_{\mu\nu}={\rm diag}~(-1,1,1,1) $ is the usual Minkowski flat metric.

To study the stress stablility criterion of the gauged Q-ball
we restrict attention to to solutions of the model \re{lag-Coleman} with spherical symmetry. The standard ansatz
for the fields is
\be
\phi(\vec r, t)=X(r)e^{i\omega t}\, ;\qquad  \,
A_0(\vec r, t)= A(r)\, , \quad A_k(\vec r, t)= 0
\label{ansatz-1}
\ee
where $X(r)$ and $A(r)$ are real functions.
Substituting it into the field equations \re{colemal_fe},\re{max}, we arrive at
\be
\begin{split}
X^{\prime\prime} + \frac{2 X^\prime}{r} + (\omega+g A)^2 X - aX
+ 2b X^3 - 3X^5 &=0\,;\\
A^{\prime\prime}+\frac{2 A^\prime}{r}
-2g(\omega+g A) X^2 &=0 \, ,
\label{eqs-ans}
\end{split}
\ee
where the prime denotes differentiation with respect to radial coordinate.

Note that  properties of a Q-ball are are qualitatively similar to those of a droplet of a liquid \cite{Mai:2012yc}.
Indeed, the energy density of the gauged Q-ball is given by
\be
\varepsilon=T_{00}=\frac12 (A^\prime)^2 + (X^\prime)^2 + (\omega+g A)^2 X^2 +
aX^2 - b X^4 + X^6
\label{eng}
\ee
It was pointed out  \cite{Polyakov:2002yz,Mai:2012yc} that the stress tensor $T_{ij}$ can be decomposed
as
\be
T_{ij} = (\hat r_i \hat r_j-\frac{1}{3}\delta_{ij})s(r)+\delta_{ij}p(r) \, ,
\label{stress-coleman}
\ee
where its trace is associated with the radial distribution of
the pressure $p(r)$ inside the Q-ball and it traceless part yields the pressure anisotropy (shear forces) $s(r)$:
\be
p(r)=(\omega+g A)^2 X^2-\frac{1}{3}(X^\prime)^2 + \frac{1}{6}(A^\prime)^2 -aX^2 + b X^4 - X^6 \, ; \qquad
s(r)= 2 (X^\prime)^2 - (A^\prime)^2
\label{p-s-coleman}
\ee
Since the  stress tensor \re{stress-coleman} is conserved,
there is a differential relation between the functions $p(r)$ and $s(r)$ \cite{Mai:2012yc,Loginov:2020xoj}
\be
d(r)=\frac{2}{r}s(r)+\frac{2}{3}s^\prime(r)+p^\prime(r)=0
\ee
Indeed, substituting the definitions \re{p-s-coleman} into this equation, we obtain
\be
\begin{split}
d(r)&=A^\prime\left(-A^{\prime\prime} -\frac{2 A^\prime}{r}
+2g(\omega+g A) X^2 \right)\\
&+ 2 X^\prime\left(X^{\prime\prime} + \frac{2 X^\prime}{r} +
(\omega+g A)^2 X - aX + 2b X^3 - 3X^5 \right)=0 \, ,
\end{split}
\ee
due to the field equations \re{eqs-ans}.
Another restriction is the von Laue condition \cite{Laue:1911lrk,Bialynicki-Birula:1993shm},
related with the internal forces balance inside a Q-ball
\be
\int\limits_0^\infty dr~r^2 p(r)=0 \, .
\label{Laue}
\ee
As a consequence, the pressure function $p(r)$ must possess an  least one zero.
This is a necessary (though not sufficient) condition for stability of the
configuration, it can be reformulated as a virial relation for the gauged Q-ball.
The condition \re{Laue} is satisfied for all solutions that will be discussed below,
it secures the stability against collapse \cite{Laue:1911lrk,Bialynicki-Birula:1993shm}.

In order to prove the von Laue condition \re{Laue}, we can integrate it by parts
imposing the finite upper integration limit $R$ \cite{Mai:2012yc}:
\be
\int\limits_0^R dr~r^2 p(r) = \left[\frac{r^3}{3}p(r) \right]_0^R - \int\limits_0^R dr~\frac{r^3}{3} p^\prime(r)
\ee
Further, since $p^\prime =-\frac{2}{3}s^\prime(r) - \frac{2}{r}s(r)$,
using the definition of the shear force \re{p-s-coleman}, we obtain
\be
p^\prime(r) = -\frac{4}{3r^3}\left[r^3 (X^\prime)^2 \right]^\prime + \frac{2}{3r^3}
\left[r^3 (A^\prime)^2 \right]^\prime = -\frac{2}{3r^3}\left[r^3 (s(r))^2 \right]^\prime
\ee
and, similar to the case of ungauged Q-balls \cite{Mai:2012yc}
\be
\int\limits_0^R dr~r^2 p(r) = \left[\frac{r^3}{3}\left(p(r) +\frac23 s(r) \right) \right]_0^R \, .
\label{int-Laue}
\ee
The asymptotic behavior of the solutions of the system \re{eqs-ans} secures the vanishing of this integral in the
limit $R\to \infty$.

The expression in brackets in \re{int-Laue} corresponds to the distribution of the normal component of
the  net force acting on an infinitesimal area element $dA \hat r_i$
at a distance $r$, is  \cite{Perevalova:2016dln,Polyakov:2018zvc}
$$
F_i(r)= dA T_{ij} \hat r^j = \left(p(r)+ \frac23 s(r) \right) dA \hat r_i \, .
$$
The corresponding  stronger local stability criterion is that the normal force must be directed outwards,
i.e.\footnote{It was conjectured that this condition can be reformulated as a restriction imposed on the longitudinal
component of the speed of sound $v_l$  \cite{Polyakov:2018zvc,Polyakov:2018rew}, its square remains positive as the $C(r)>0$.
However, the validity of this approximation merits further study.}
\be
C(r)=\frac23 s(r) + p(r) > 0
\label{criterion}
\ee

We will see that this condition is not always satisfied for the $U(1)$ gauged Q-balls.

\subsubsection{Numerical results}

To find numerical solutions of the coupled partial differential equations \re{eqs-ans},
we implement the fourth-order finite-difference method. The system of equations is discretized on an equidistant grid
in radial coordinate $r$. We map the infinite interval of the variable $r$ onto the compact radial coordinate
$x=\frac{r/r_0}{1+r/r_0}  \in [0:1]$. Here $r_0$ is a real scaling constant, which typically is taken as $r_0 = 2 - 15$.
We impose the following set of the boundary conditions:
\be
\partial_r X(0)=0, \quad \partial_r A(0)=0,
\label{bc-six-a}
\ee
and
\be
X(\infty)=0, \quad A(\infty)=0.
\label{bc-six-b}
\ee
As usual, they follow from conditions of regularity of the fields at the origin, from the definition of the vacuum
at spatial infinity and our choice of the gauge.

There are some important differences between the  gauged and ungauged Q-balls
\cite{Lee:1988ag,Anagnostopoulos:2001dh,Gulamov:2013cra,Gulamov:2015fya,Nugaev:2019vru,Loiko:2019gwk}.
Both the energy and the charge of the global ($g=0$) Q-balls diverge as the angular frequency approaches the
critical values $\{\omega_{min},\omega_{max},\} $. On the contrary, the gauged Q-balls possess finite energy and
charge for all range of values of the angular frequency. The gauged Q-balls form a first (lower in energy) branch
of solutions which extends backward as $\omega$ decreases below the maximal value, see Fig.\ref{fig1}, upper left plot.
\begin{figure}[h!]
\begin{center}
\includegraphics[height=.32\textheight,  angle =-90]{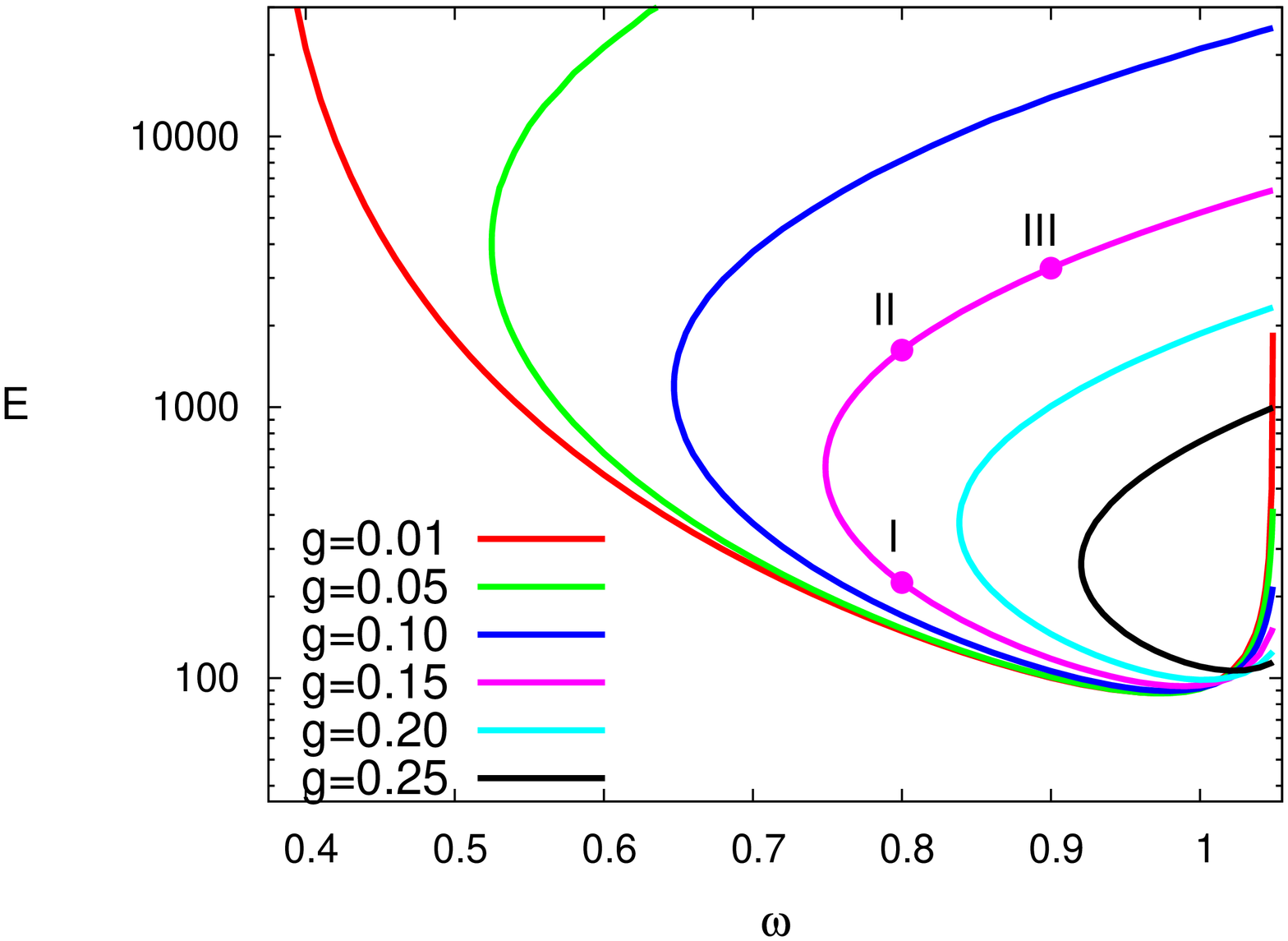}
\includegraphics[height=.32\textheight,  angle =-90]{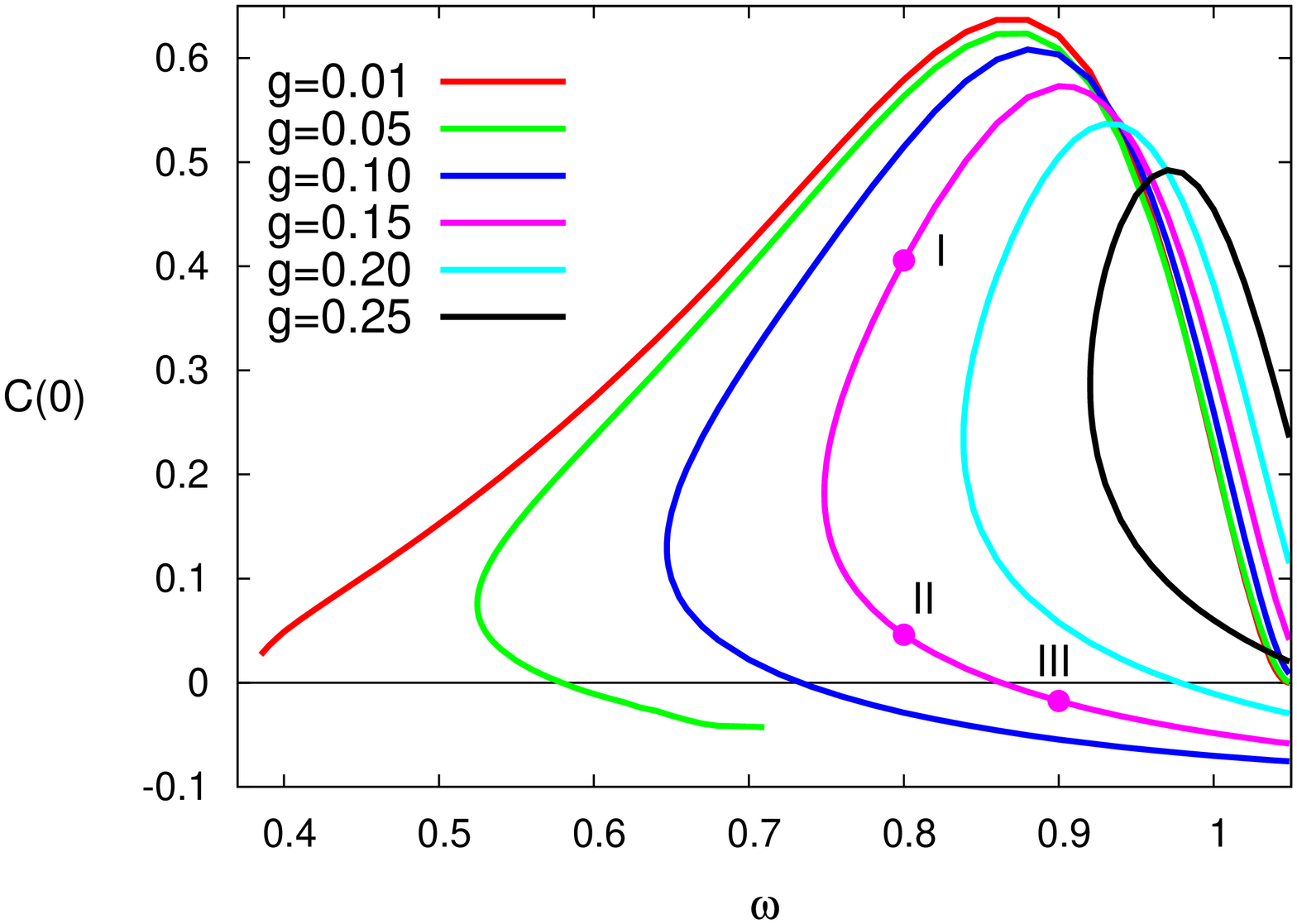}
\includegraphics[height=.32\textheight,  angle =-90]{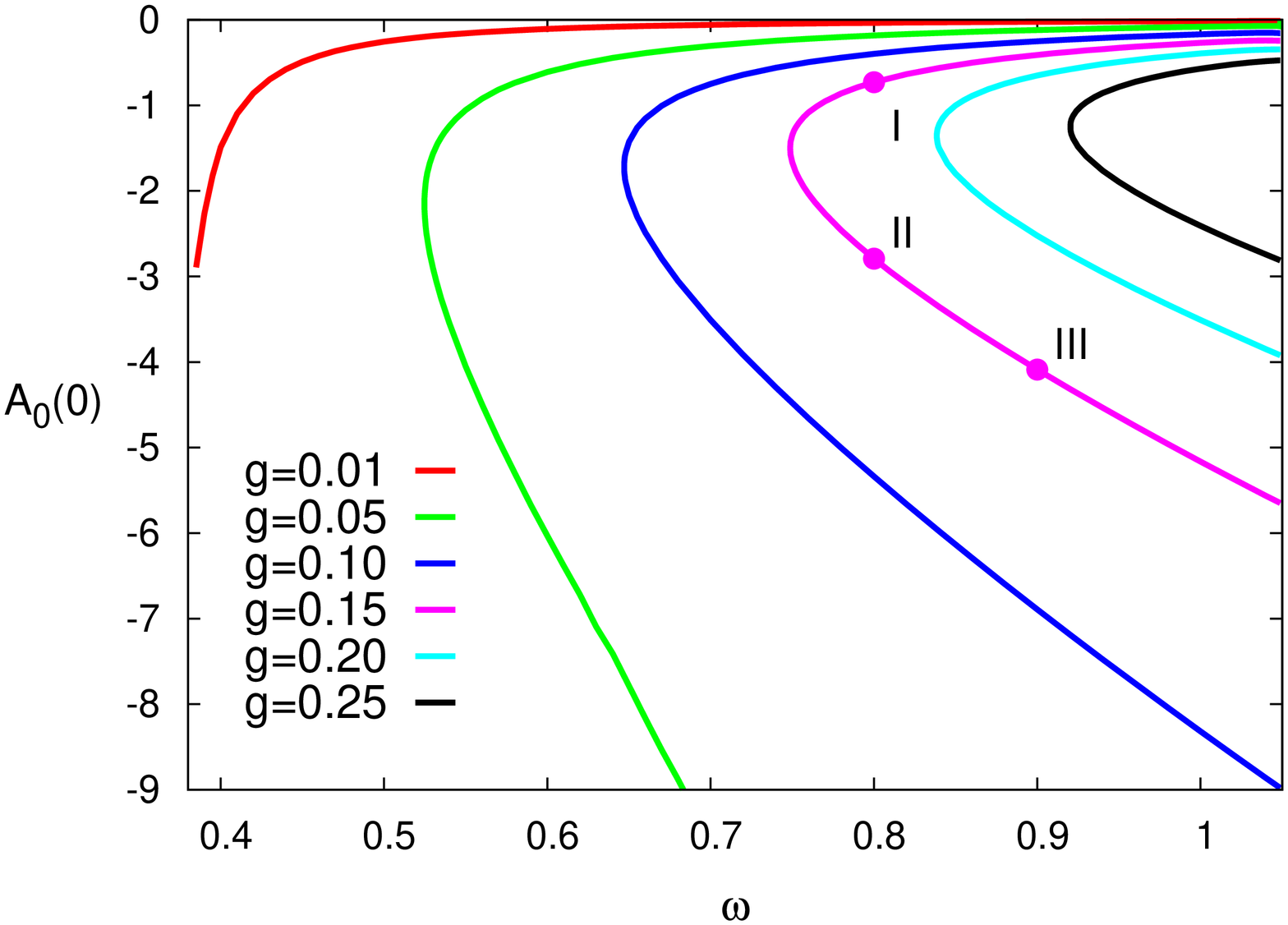}
\includegraphics[height=.32\textheight,  angle =-90]{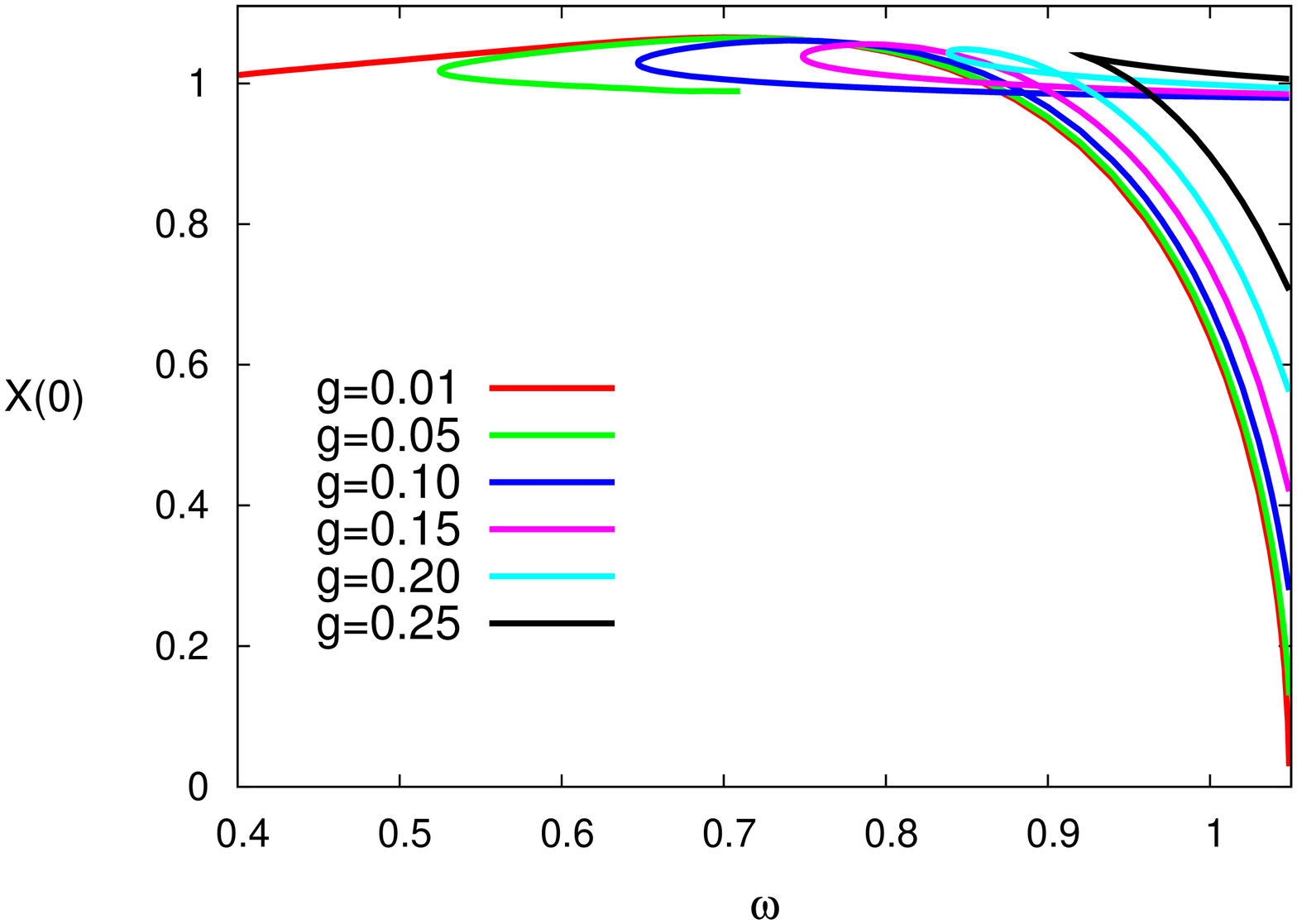}
\end{center}
\caption{\small
$U(1)$ gauged Q-balls in the model \re{lag-Coleman}: The total energy of the solutions (upper left plot),
the values of the function $C(0)$ (upper right plot), the gauge
potential $A_0(0)$ (bottom left plot) and the scalar profile function $X(0)$ (bottom right plot) at $r=0$ are displayed as
functions of the angular frequency $\omega$ for some set of values of the gauge coupling $g$.}
    \lbfig{fig1}
\end{figure}
Along this branch the electrostatic energy of the configuration remains much smaller than the total energy
of the Q-ball, the solutions are not very different from the global Q-balls. The size of the soliton
increases as the angular frequency decreases, similar effect is observed as  the gauge coupling grows, see
Fig.~\ref{fig3}, upper left plot. Note that on the first branch the shear force function $s(r)$ possess a
maximum associated with the node of the pressure $p(r)$, as displayed in  Fig.~\ref{fig3}. The pressure distribution
is positive in the core of the Q-ball and negative in the outer region.
The corresponding function $C(r)$ \re{criterion} is positive, on the lower (scalar)
branch the gauged Q-balls  are stable with respect to internal deformations.

The second branch of gauged Q-balls is formed
at the minimal critical value $\omega_{min}$, this branch extends forward as $\omega$ increases
again towards the upper critical value $\omega_{max}$.
The energy of the electrostatic repulsion begin dominating over the scalar interaction,
when the bifurcation with the second higher energy branch is approached. Along the upper
branch the characteristic size of the gauged Q-balls continues to increase,
the strong electrostatic repulsive force inflates the configuration.
The lower critical value $\omega_{min}$
is increasing as the gauge coupling $g$ becomes larger.

Note that the classical ungauged  Q-ball is classically stable if \cite{Panin:2016ooo}
\be
\label{cond-stable}
\frac{\omega}{Q} \frac{d Q}{d \omega} < 0 \, .
\ee
This is so-called Vakhitov-Kolokolov criteria of stability
\cite{Vakhitov}. It was argued that this criteria cannot be applied to gauged Q-balls \cite{Panin:2016ooo}.
As we  see from the upper left plot in the Fig.~\ref{fig3}, the  Vakhitov-Kolokolov  inequality \re{cond-stable}
does not hold  for the gauged Q-balls  on the upper branch. Further, the criteria \re{criterion} is violated for such solutions.
Indeed, in Fig.~\ref{fig4} we displayed the profile function of the scalar field $X(r)$,
distributions of the pressure $p(r)$, and the shear forces $s(r)$ \re{p-s-coleman} and the
criteria function $C(r)$ \re{criterion} for the gauged Q-balls on the second electrostatic branch.  Further, to
investigate the pattern of evolution of the gauged Q-balls, in Fig.~\ref{fig2} we presented the distributions of the
total energy density $\varepsilon(r)$ \re{eng} and the functions $p(r)$, $s(r)$ and
$C(r)$ of the particular solutions, labeled as $I,II$ and $III$ on the Fig~\ref{fig1}.

Comparing these solutions and the corresponding plots shown in  Figs.~\ref{fig3},\ref{fig4}, we can clearly see that the
stability criteria
\re{criterion} is violated on the second branch. Further, the corresponding pressure function $p(r)$
possess more than one node while the shear force  $s(r)$ becomes negative both inside of the core of
the configuration and on the spacial asymptotic. In other words, electrostatic repulsion tears the gauged Q-ball apart.

Note that the electrostatic energy depends both on the angular frequency $\omega$ and on the value of the gauge coupling $g$.
For a fixed value of $g$ its contribution to the total energy increases as $\omega$ decreases, while for a fixed value of $\omega$
it increases as $g$ becomes larger. Consequently, as the coupling $g$ remains  relatively small,
$g\lesssim g_{cr}\approx  0.07$, at some critical values of the angular frequency the function $C(r)$ may
become negative everywhere in space, see Fig.~\ref{fig4}, bottom right plot. Indeed, our numerical scheme fails
to find a second branch solution as the gauge coupling decreases below a critical value $g_{cr}$.

\begin{figure}[h!]
\begin{center}
\includegraphics[height=.32\textheight,  angle =-90]{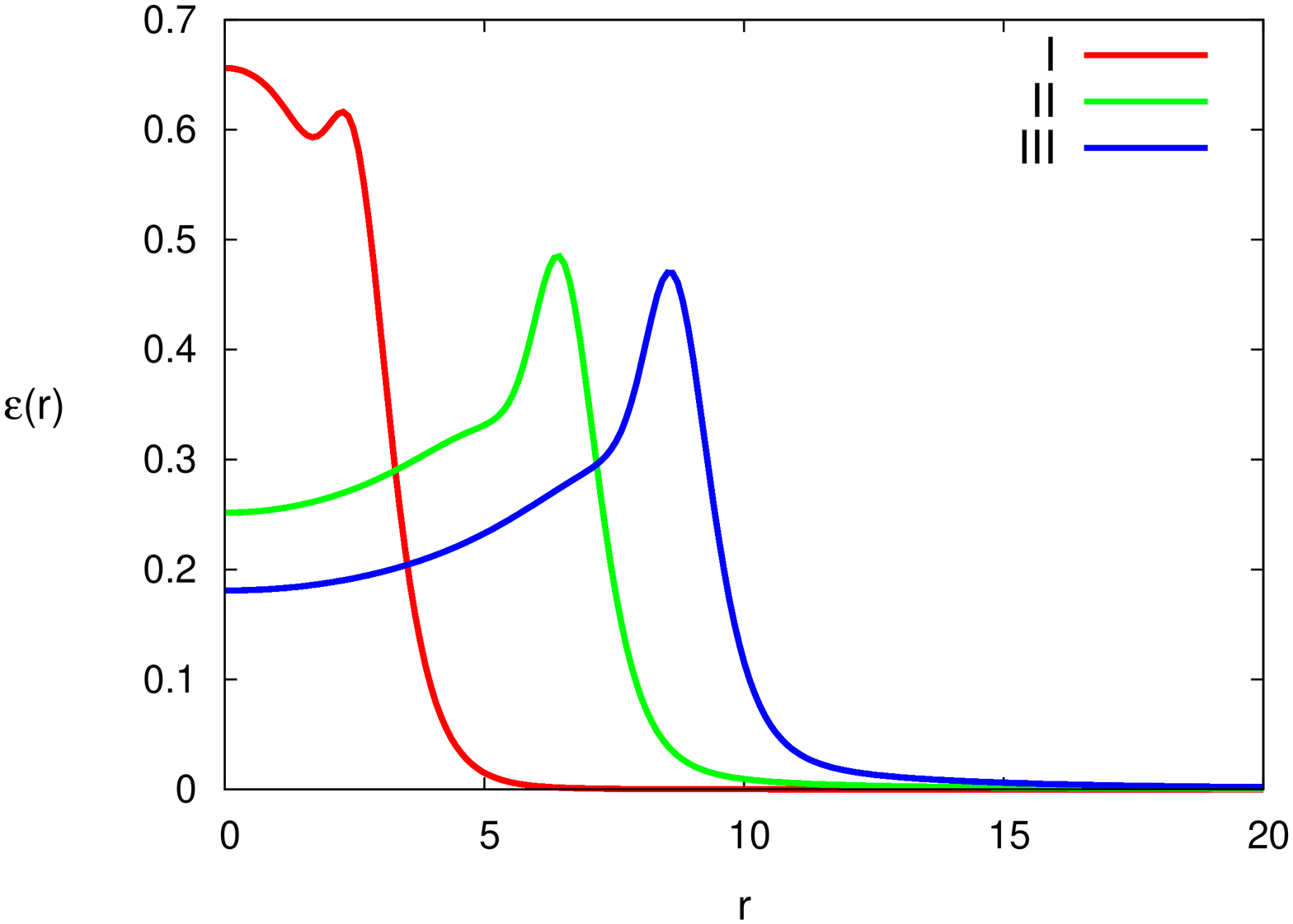}
\includegraphics[height=.32\textheight,  angle =-90]{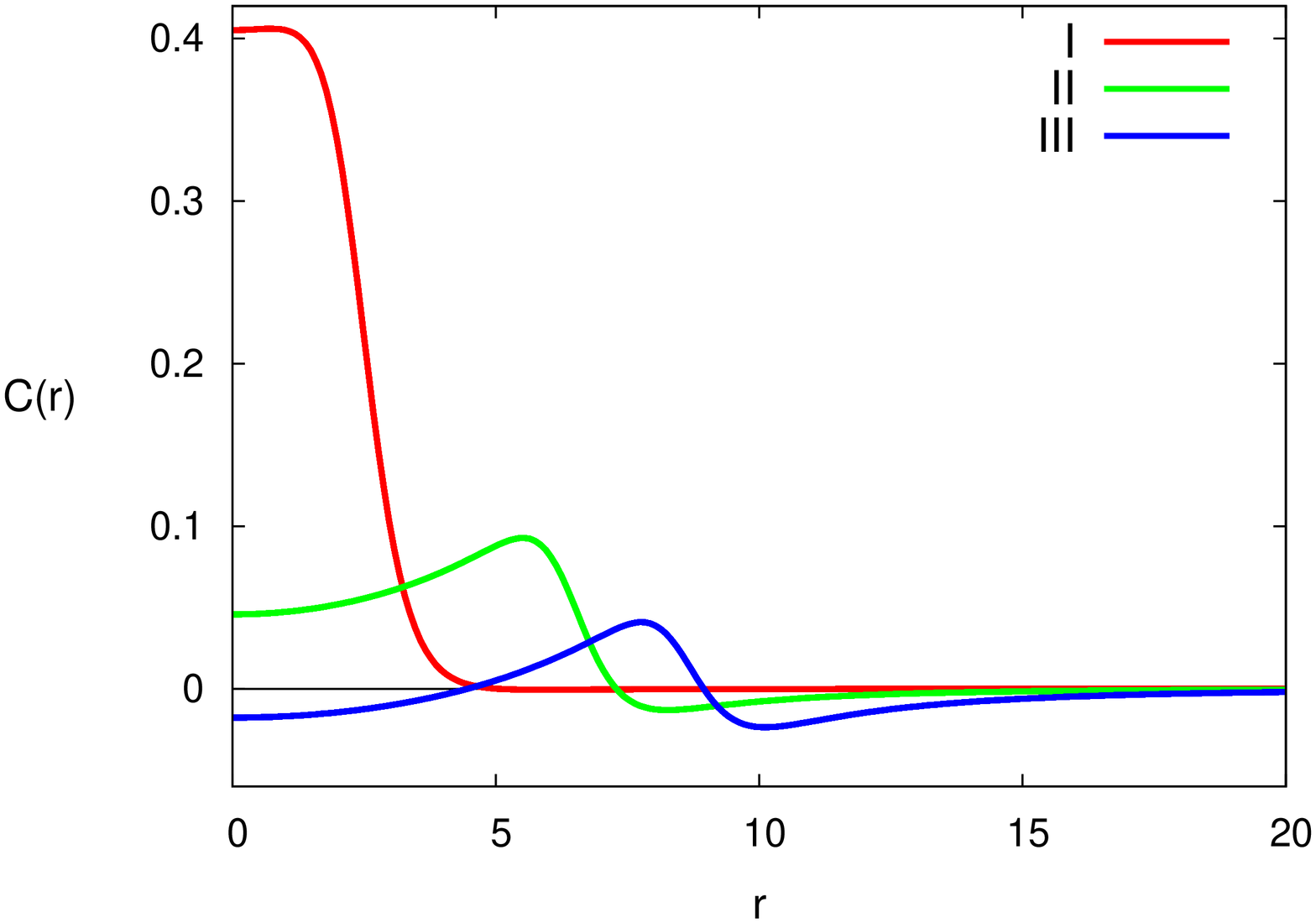}
\includegraphics[height=.32\textheight,  angle =-90]{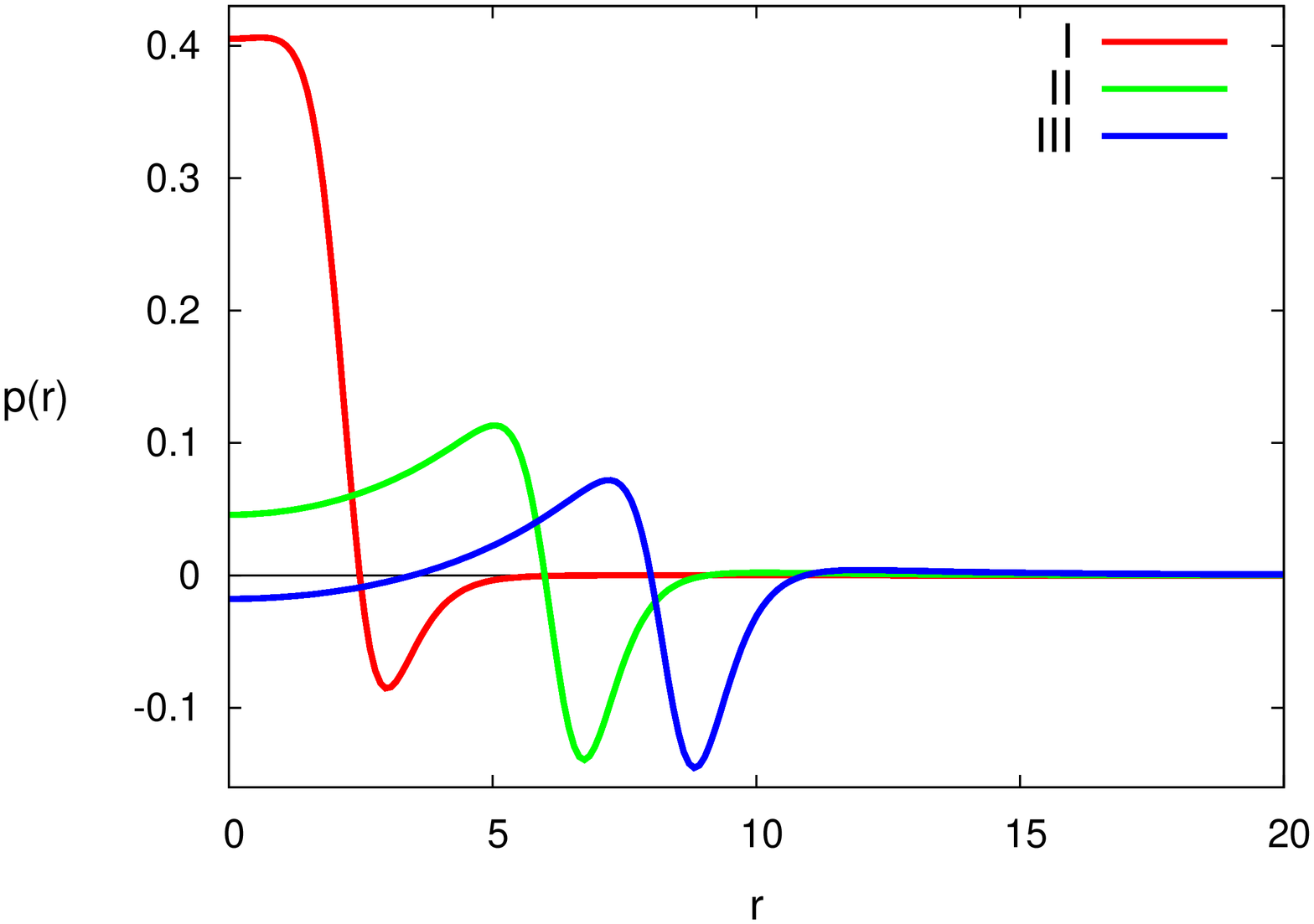}
\includegraphics[height=.32\textheight,  angle =-90]{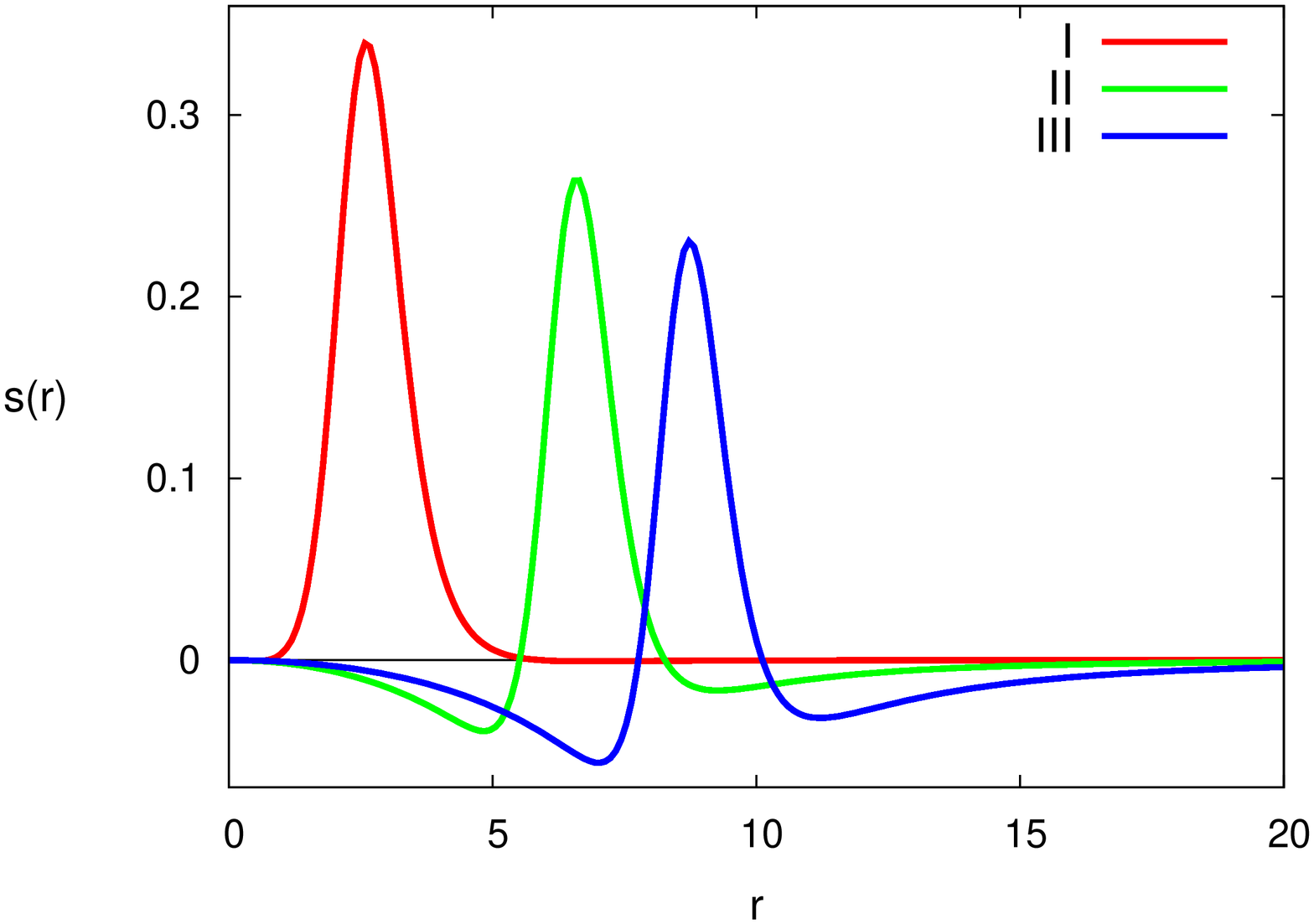}
\end{center}
\caption{\small
Solutions of the model \re{lag-Coleman}, labelled as points $I,II$ and $III$ on the Fig.\ref{fig1}: The distributions of the  energy density $\varepsilon(r)$  (upper left plot), the function $C(r)$ (upper right plot), the pressure function  $p(r)$ (bottom left plot) and the shear force function $s(r)$ (bottom right plot).}
    \lbfig{fig2}
\end{figure}

\begin{figure}[h!]
\begin{center}
\includegraphics[height=.32\textheight,  angle =-90]{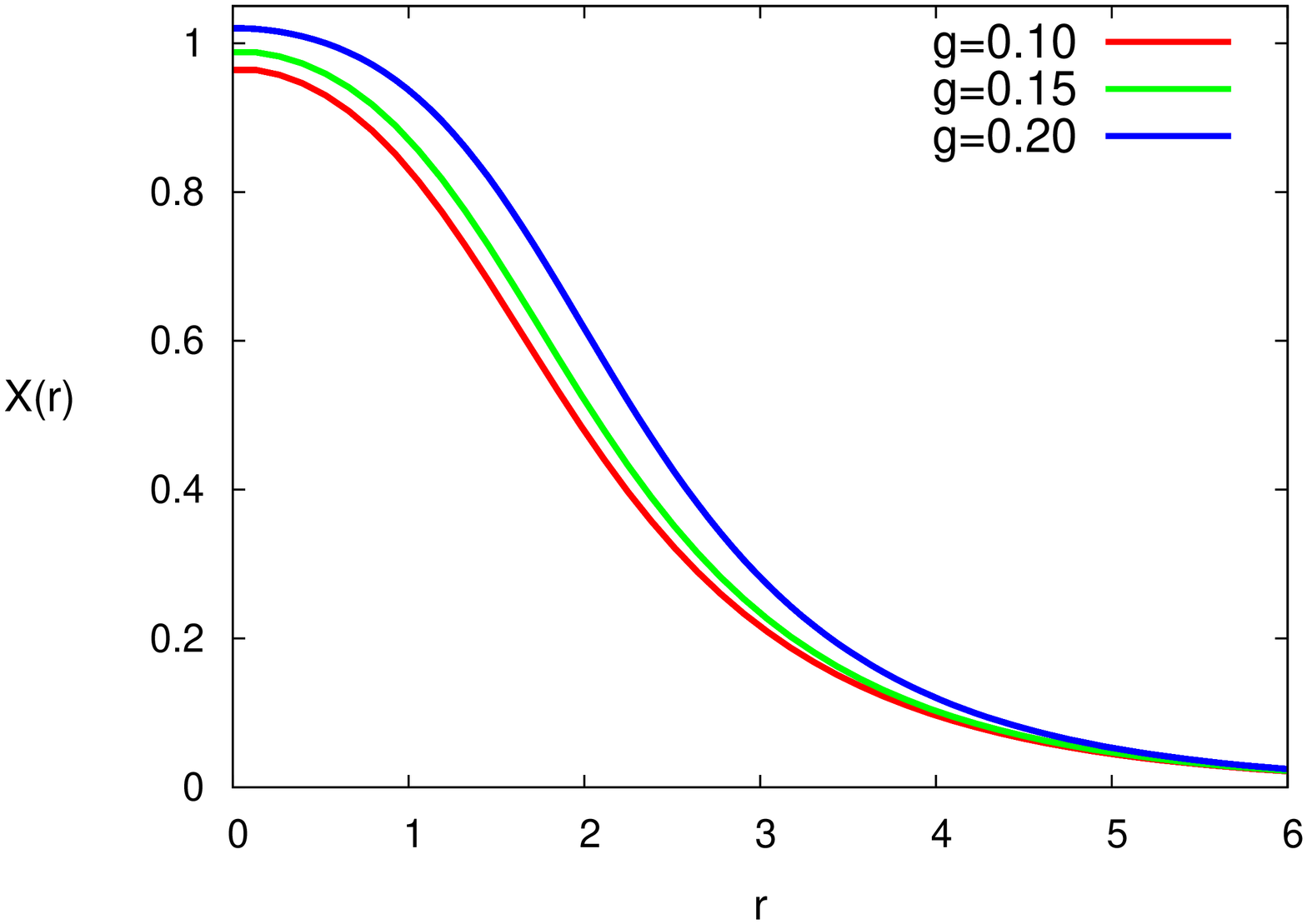}
\includegraphics[height=.32\textheight,  angle =-90]{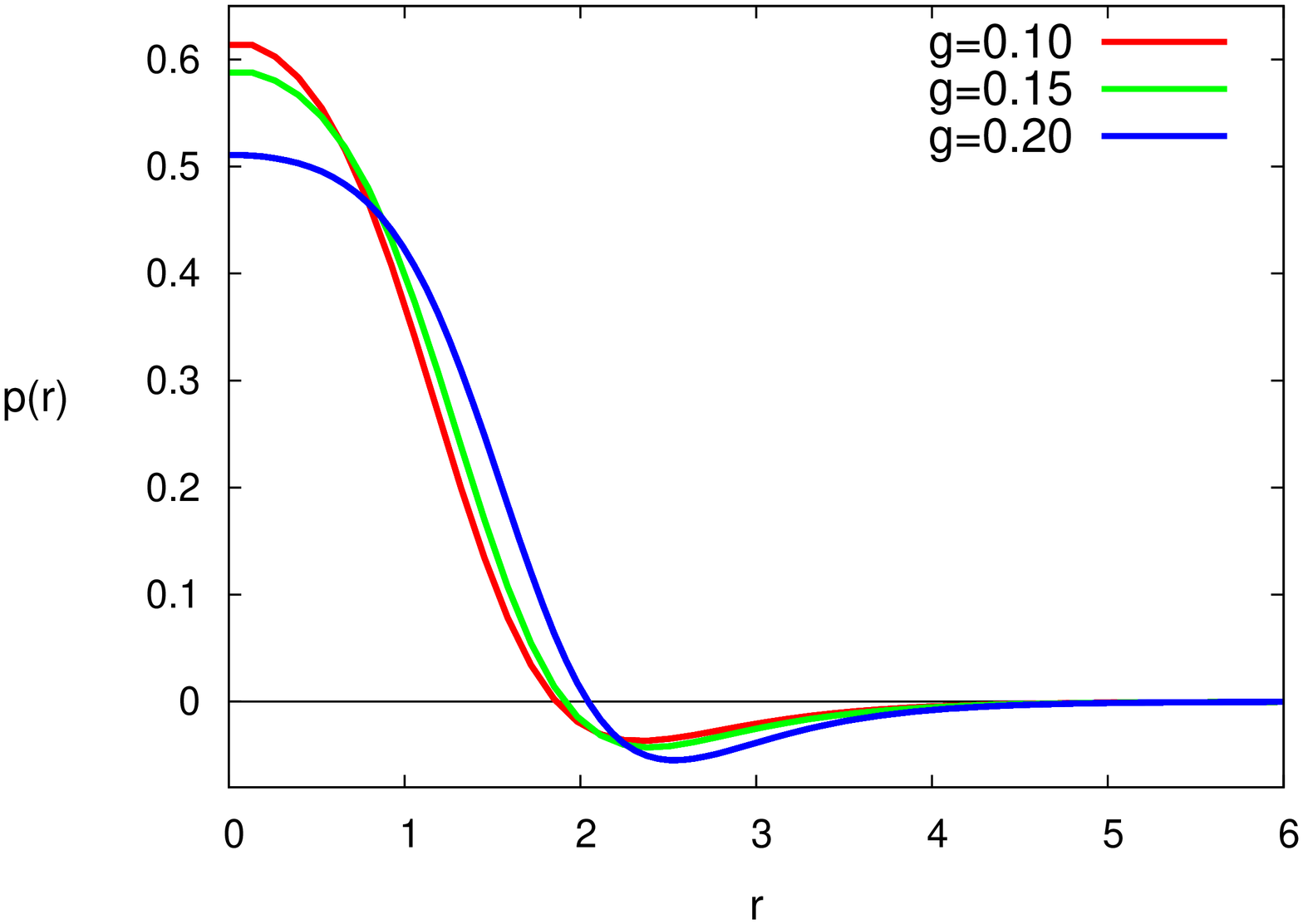}
\includegraphics[height=.32\textheight,  angle =-90]{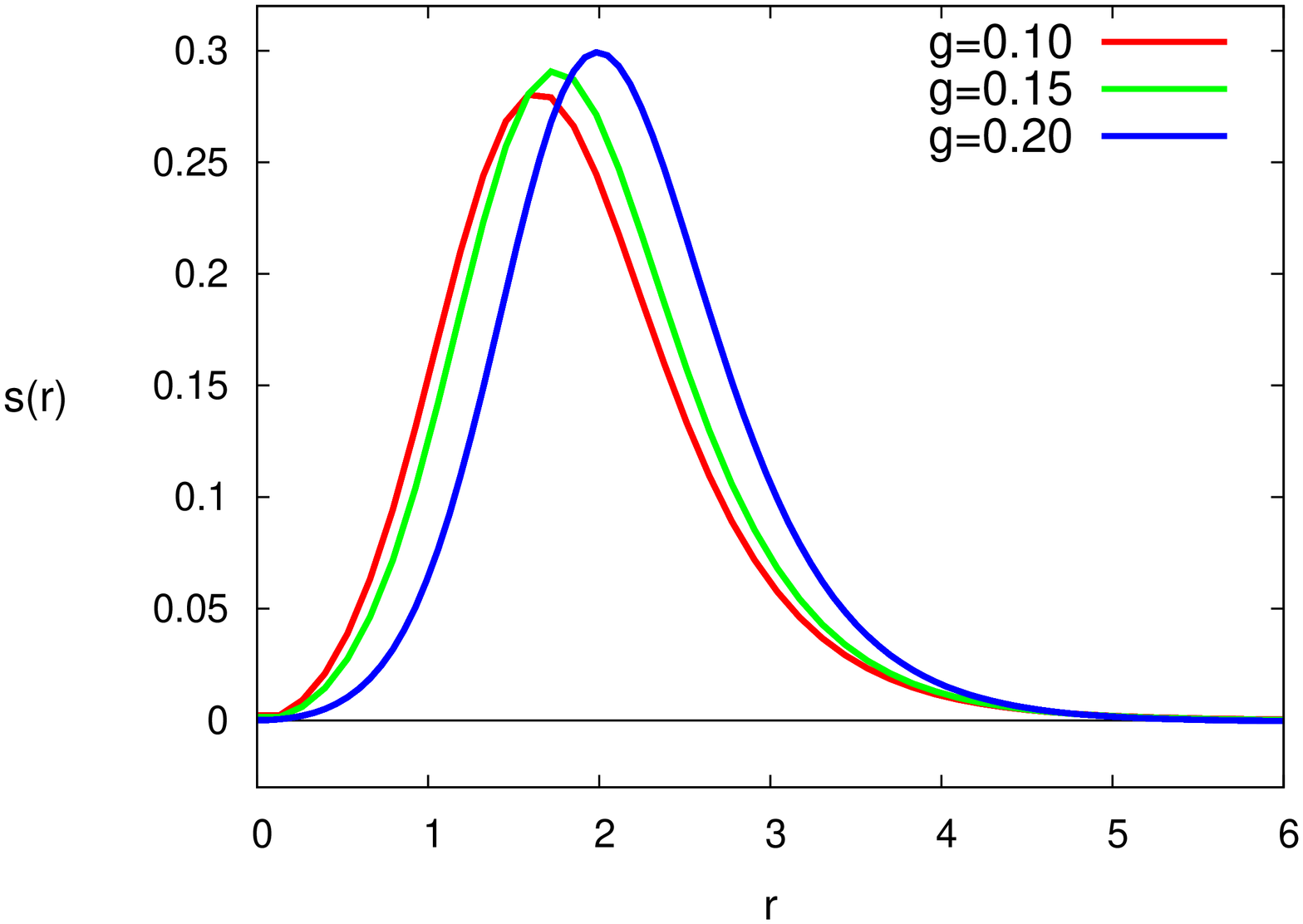}
\includegraphics[height=.32\textheight,  angle =-90]{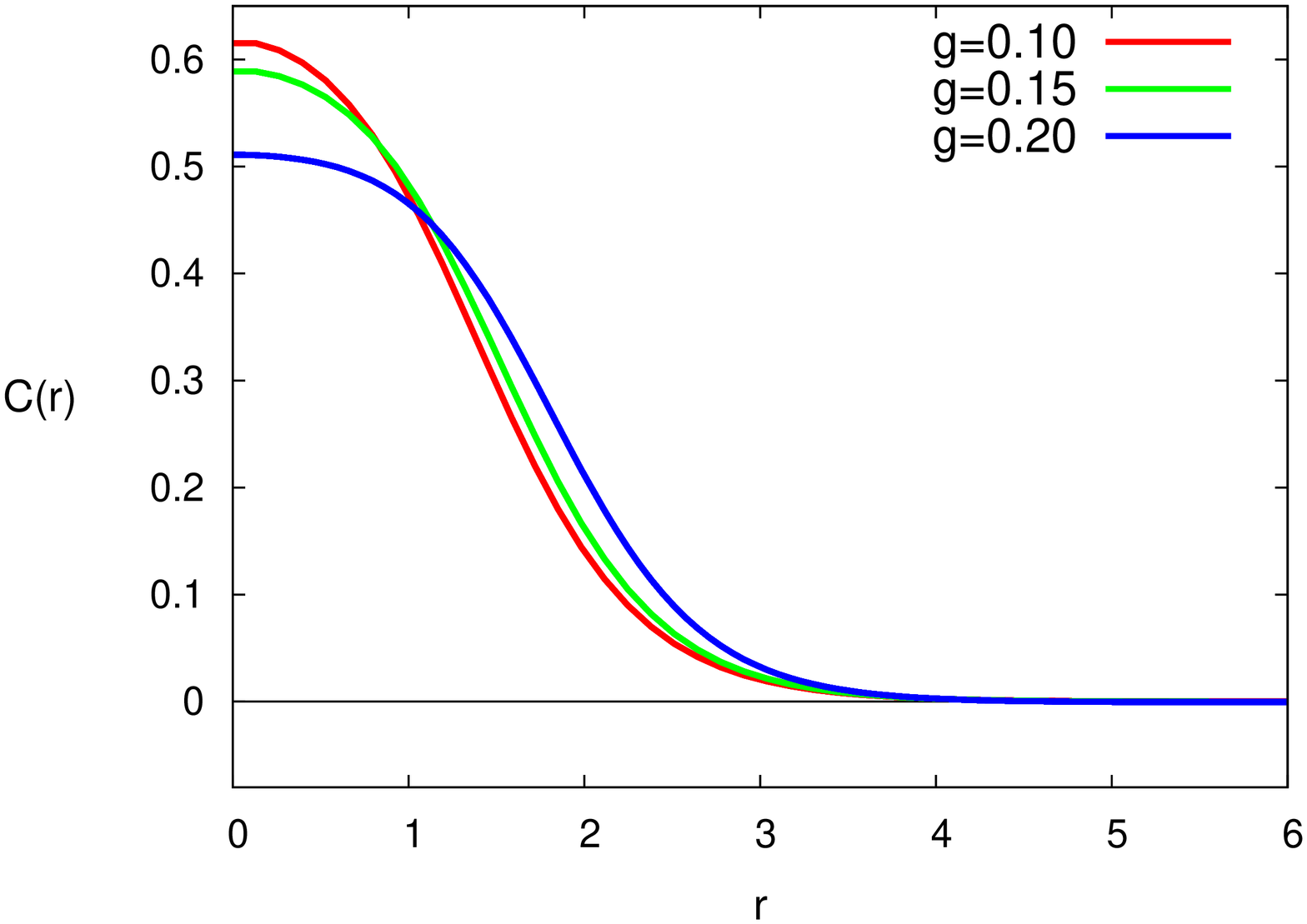}
\end{center}
\caption{\small
$U(1)$ gauged Q-balls in the model \re{lag-Coleman} on the lower (scalar) branch: The profile function of the scalar field $X(r)$, the pressure $p(r)$, the shear force $s(r)$ and the function $C(r)$ are displayed at $\omega=0.90$
for some set of  values of the gauge coupling $g$.}
    \lbfig{fig3}
\end{figure}

\begin{figure}[h!]
\begin{center}
\includegraphics[height=.32\textheight,  angle =-90]{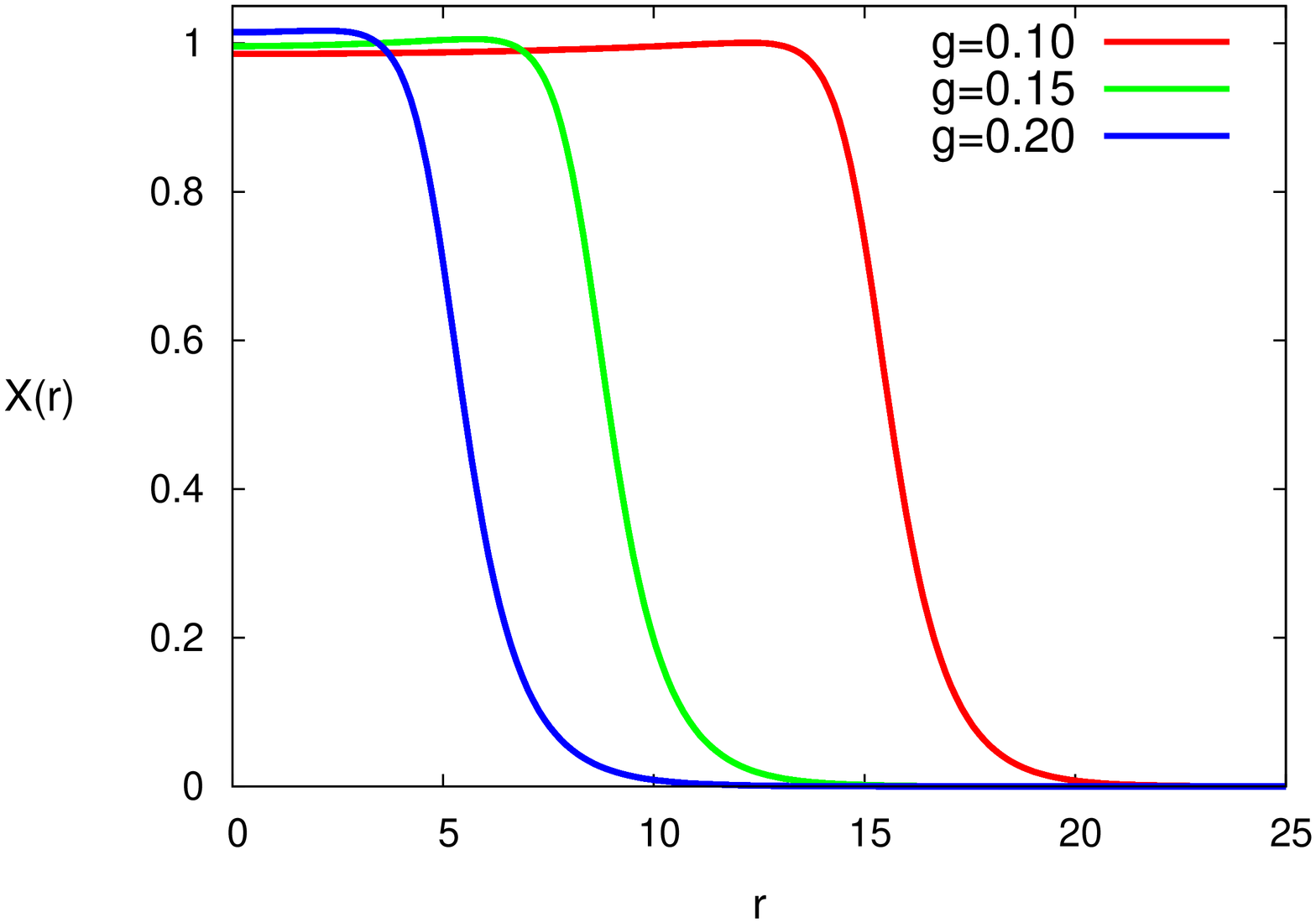}
\includegraphics[height=.32\textheight,  angle =-90]{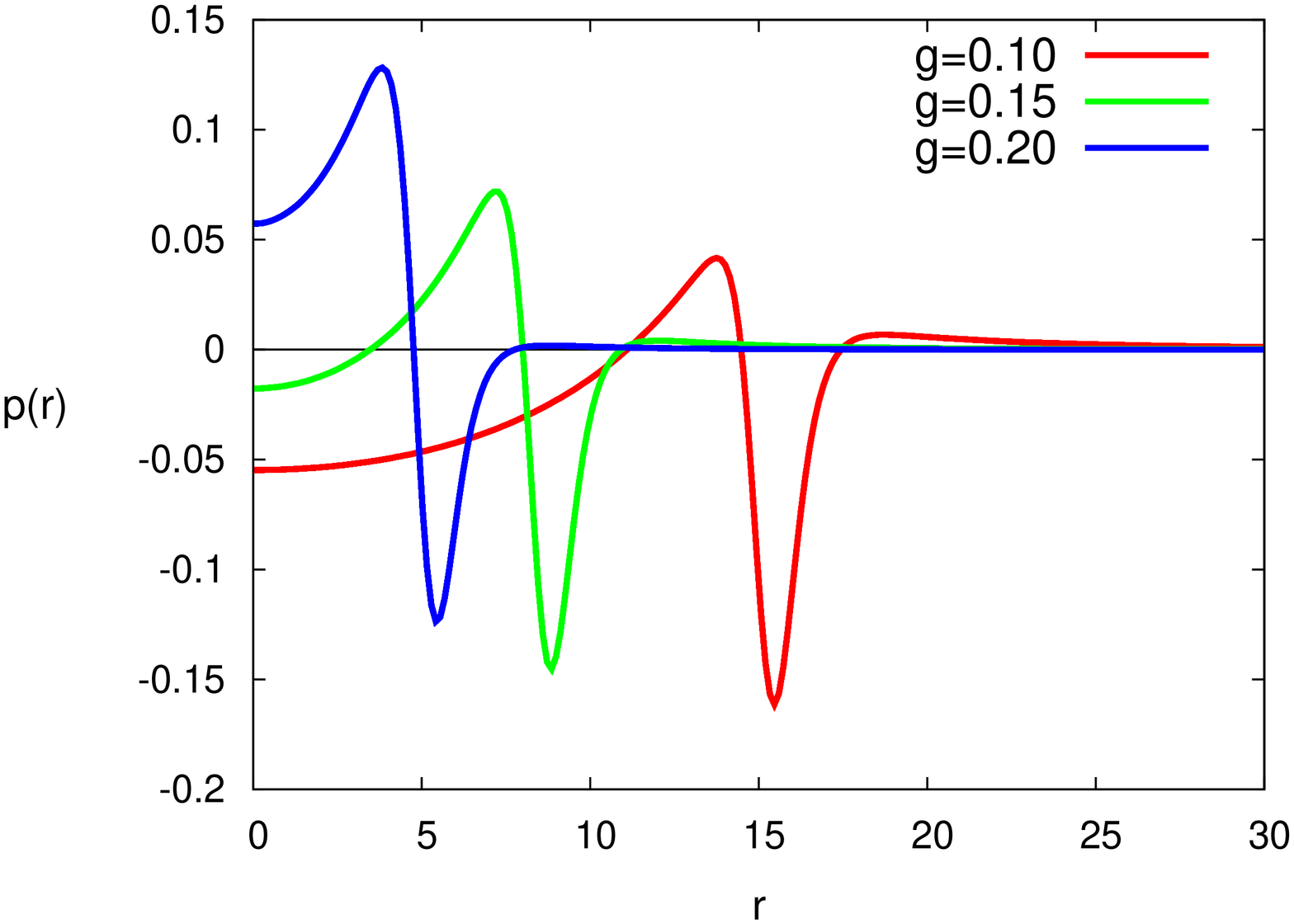}
\includegraphics[height=.32\textheight,  angle =-90]{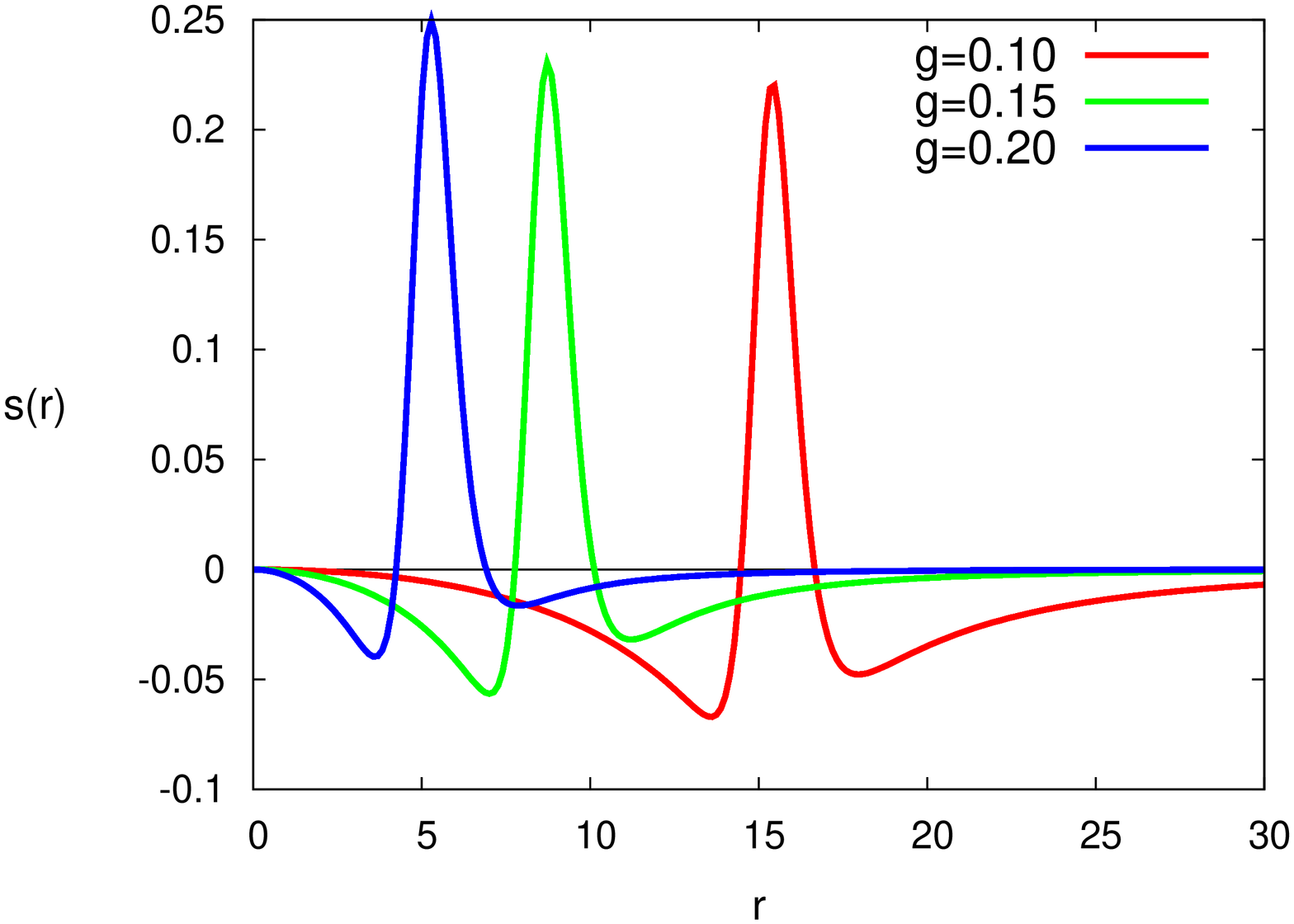}
\includegraphics[height=.32\textheight,  angle =-90]{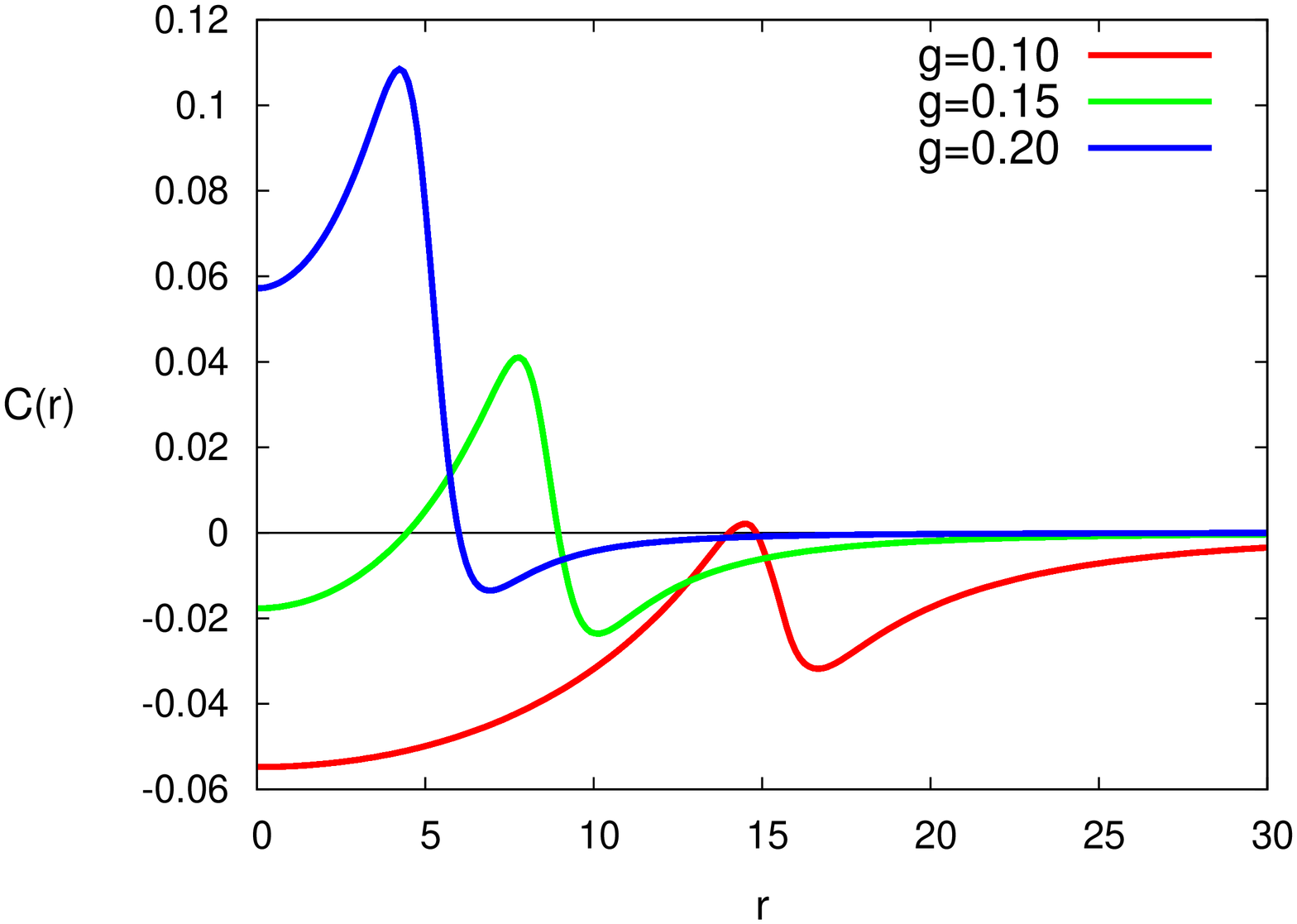}
\end{center}
\caption{\small
$U(1)$ gauged Q-balls in the model \re{lag-Coleman} on the upper (electrostatic) branch: The
profile function of the scalar field $X(r)$, the pressure $p(r)$, the shear force $s(r)$ and the
function $C(r)$ are displayed at $\omega=0.90$
for some set of values of the gauge coupling $g$.}
    \lbfig{fig4}
\end{figure}

\subsection{$U(1)$ gauged Q-balls in the two-component Friedberg-Lee-Sirlin-Maxwell model}

Another example of a model, which supports $U(1)$ gauged Q-balls, is given by the two-component Friedberg-Lee-Sirlin-Maxwell
model \cite{Lee:1991bn}. It describes a coupled system of a
real self-interacting scalar field $\psi$ with a symmetry breaking potential and a complex scalar field $\phi$, coupled to
the electromagnetic gauge potential $A_\mu$:
\be
L^{(II)}= -\frac14 F_{\mu\nu} F^{\mu\nu} + (\partial_\mu\psi)^2 + D_\mu\phi D^\mu\phi^* - m^2 \psi^2|\phi|^2 - U(\psi) \, ,
\label{lag-fls}
\ee
where $m$ is the coupling constant.  The symmetry breaking potential of the real scalar field is
\be
U(\psi)= \mu^2 (1-\psi^2)^2 \, ,
\label{pot-fls}
\ee
thus, in the vacuum $\psi \to 1$  and the complex field $\phi$ becomes massive due to the coupling
with the real component. The parameters  $\mu$ and $m$ correspond to the
mass of the real and complex components, respectively. Note that, for any finite values of the mass parameter $m$,
the complex field becomes massless when the real component is zero. Further, the gauge field $A_\mu$ acquires a mass
due to the coupling with the complex scalar field, it is long-ranged as $|\phi|=0 $.
Notably, in the limit of vanishing mass parameter $\mu \to 0$ but fixed
vacuum expectation value of the real scalar component, the field $\psi$
becomes massless and thus long-ranged.
However, the complex component $\phi$ still acquires mass in this limit
\cite{Levin:2010gp,Loiko:2018mhb}.

The system of field equations of the model \re{lag-fls} includes two coupled scalar fields equations
\be\label{fls_fe}
\begin{split}
\partial^{\mu}\partial_{\mu}\psi &= \psi\left( m^2|\phi|^2 + 2\mu^2(1-\psi^2) \right), \\
D^{\mu}D_{\mu}\phi &= m^2\psi^2\phi
\end{split}
\ee
and the Maxwell equation \re{max}.
The corresponding energy-momentum tensor reads
\be
\begin{split}
T_{\mu\nu}&=  -\eta_{\mu\nu} \left(D_\rho\phi D^\rho \phi^* + (\partial_\rho \psi)^2  + \frac14 F_{\rho\sigma} F^{\rho\sigma}
+ U(|\psi|)\right)\\
& + \left(D_\mu\phi D_\nu \phi^* + D_\mu\phi^* D_\nu \phi\right) +
\partial_\mu\psi \partial_\nu \psi
+ \eta^{\rho \sigma}
F_{\mu \rho} F_{\nu \sigma}
\end{split}
\ee

We consider the usual spherically symmetric parameterization of the fields
\be
\phi(\vec r, t)= X(r)e^{i\omega t} \, ;\qquad  \psi(\vec r, t)=Y(r) \, ,
\label{ansatz}
\ee
and $A_0(\vec r, t)= A(r),~~A_k(\vec r, t)=0$,
where $X(r), Y(r)$ and $A(r)$ are real functions of radial variable and $\omega$ is the frequency of stationary rotation.
By making use of this ansatz the system of field equations \re{fls_fe},\re{max} can be solved numerically. The boundary conditions
\re{bc-six-a},\re{bc-six-b} are extended by the following conditions on the scalar function $Y(r)$: $\partial_rY(0)=0, ~~Y(\infty)=1$.

The total energy density of the system is
(cf \re{eng})
\be
\epsilon=(X^\prime)^2+(Y^\prime)^2 + (\omega+ gA)^2 X^2 + \mu^2(1-Y^2)^2 +
m^2 X^2 Y^2 + \frac12 (A^\prime)^2 \, .
\label{eng-fls}
\ee
Making use of the decomposition of the stress tensor \re{stress-coleman}, we can identify the pressure and the shear forces of the gauged Friedberg-Lee-Sirlin-Maxwell Q-ball as
\be
\begin{split}
p(r)&=(\omega+g A)^2 X^2-\frac{1}{3}(X^\prime)^2 -\frac{1}{3}(Y^\prime)^2+ \frac{1}{6}(A^\prime)^2 -\mu^2(1-Y^2)^2 -m^2 X^2 Y^2 \, ; \\
s(r)&= 2 (X^\prime)^2 +2 (Y^\prime)^2 - (A^\prime)^2 \, .
\end{split}
\label{p-s-fls}
\ee
Hence, the Q-ball stability criteria \re{criterion} for the model \re{lag-fls} becomes
\be
C(r)=(X^\prime)^2 +(Y^\prime)^2 - \frac12(A^\prime)^2
+(\omega + gA)^2 X^2 - m^2 X^2 Y^2 - \mu^2(1-Y^2)^2 > 0
\label{cr-fls}
\ee

\begin{figure}[h!]
\begin{center}
\includegraphics[height=.33\textheight,  angle =-90]{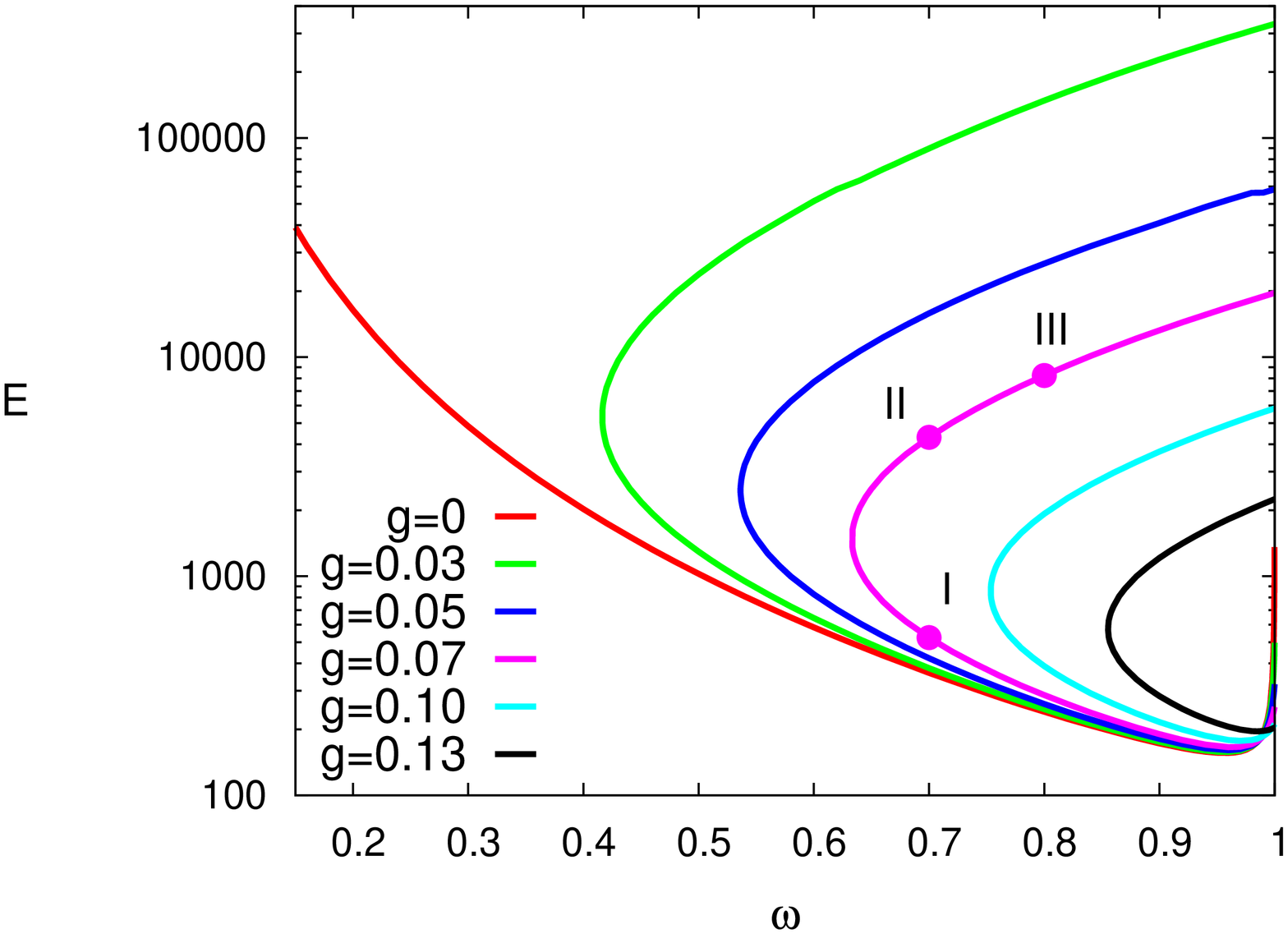}
\includegraphics[height=.33\textheight,  angle =-90]{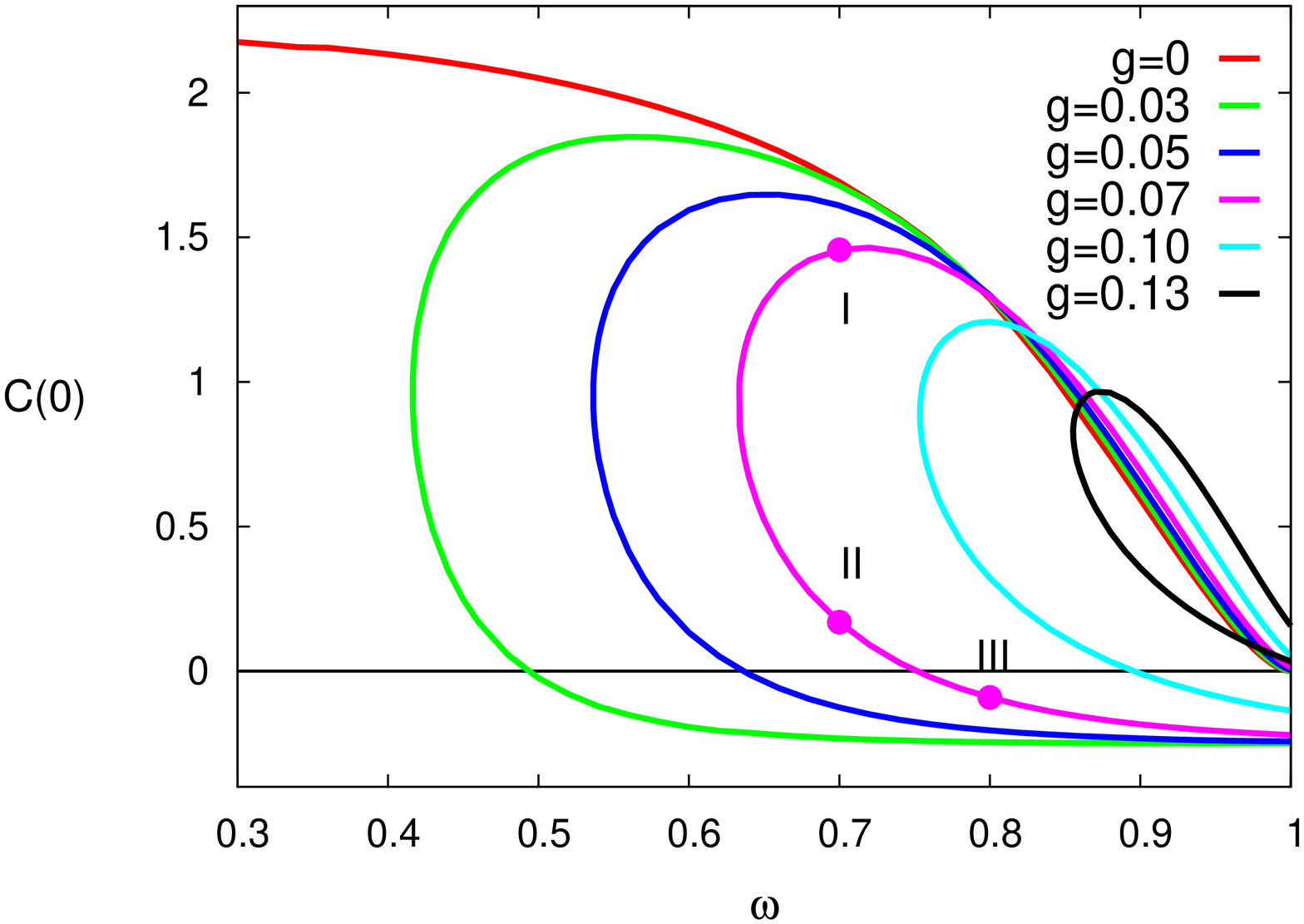}
\includegraphics[height=.24\textheight,  angle =-90]{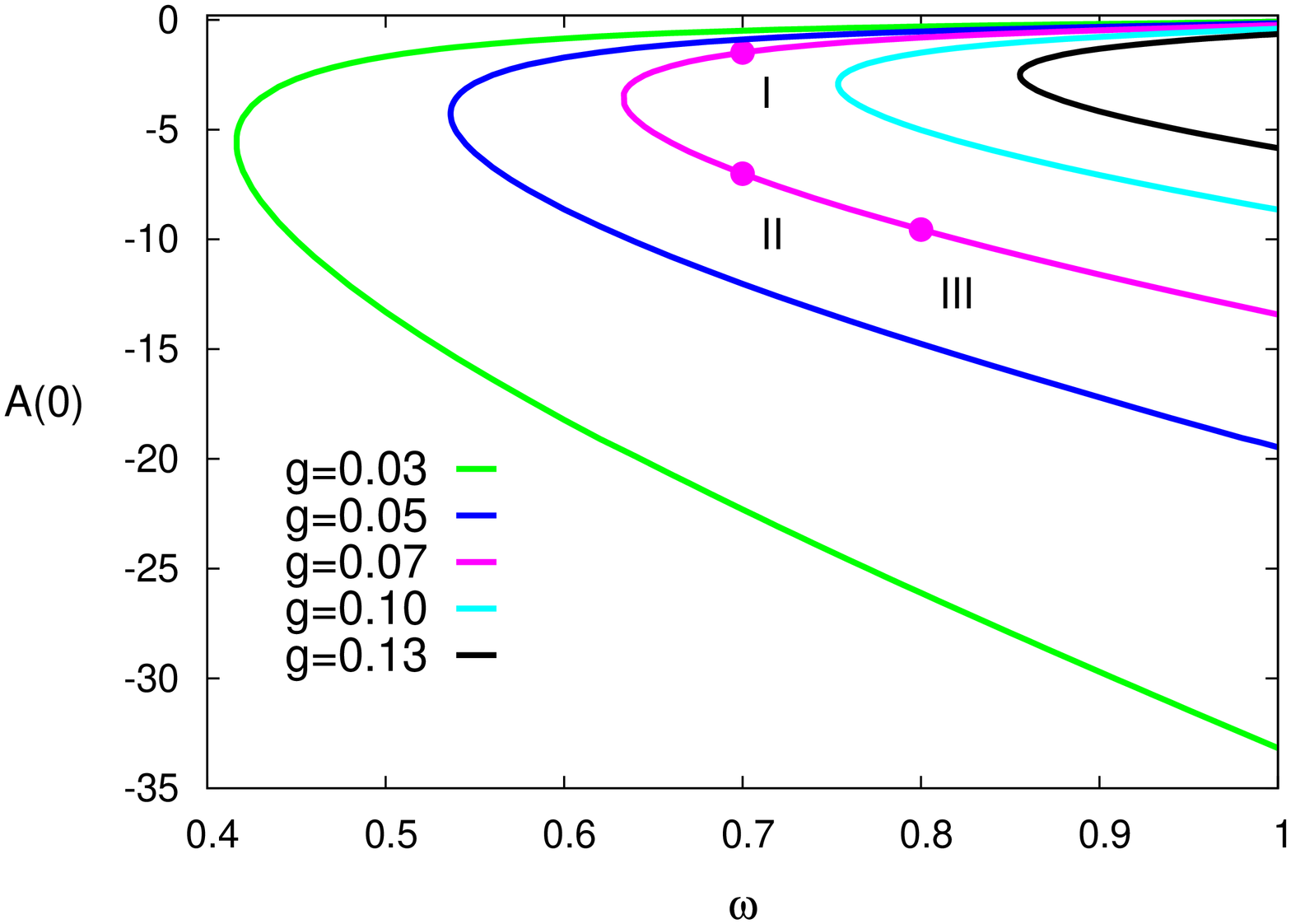}
\includegraphics[height=.24\textheight,  angle =-90]{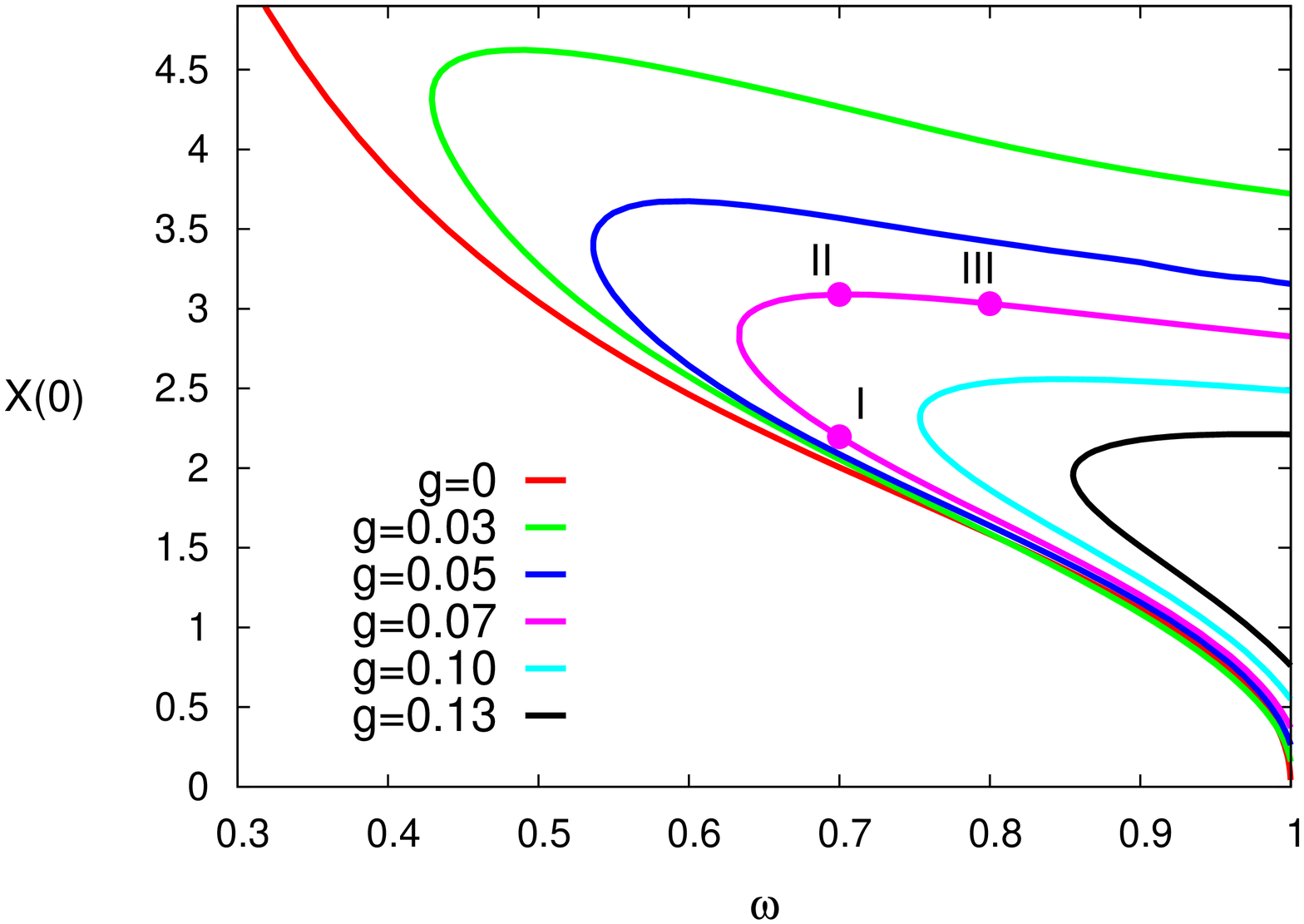}
\includegraphics[height=.24\textheight,  angle =-90]{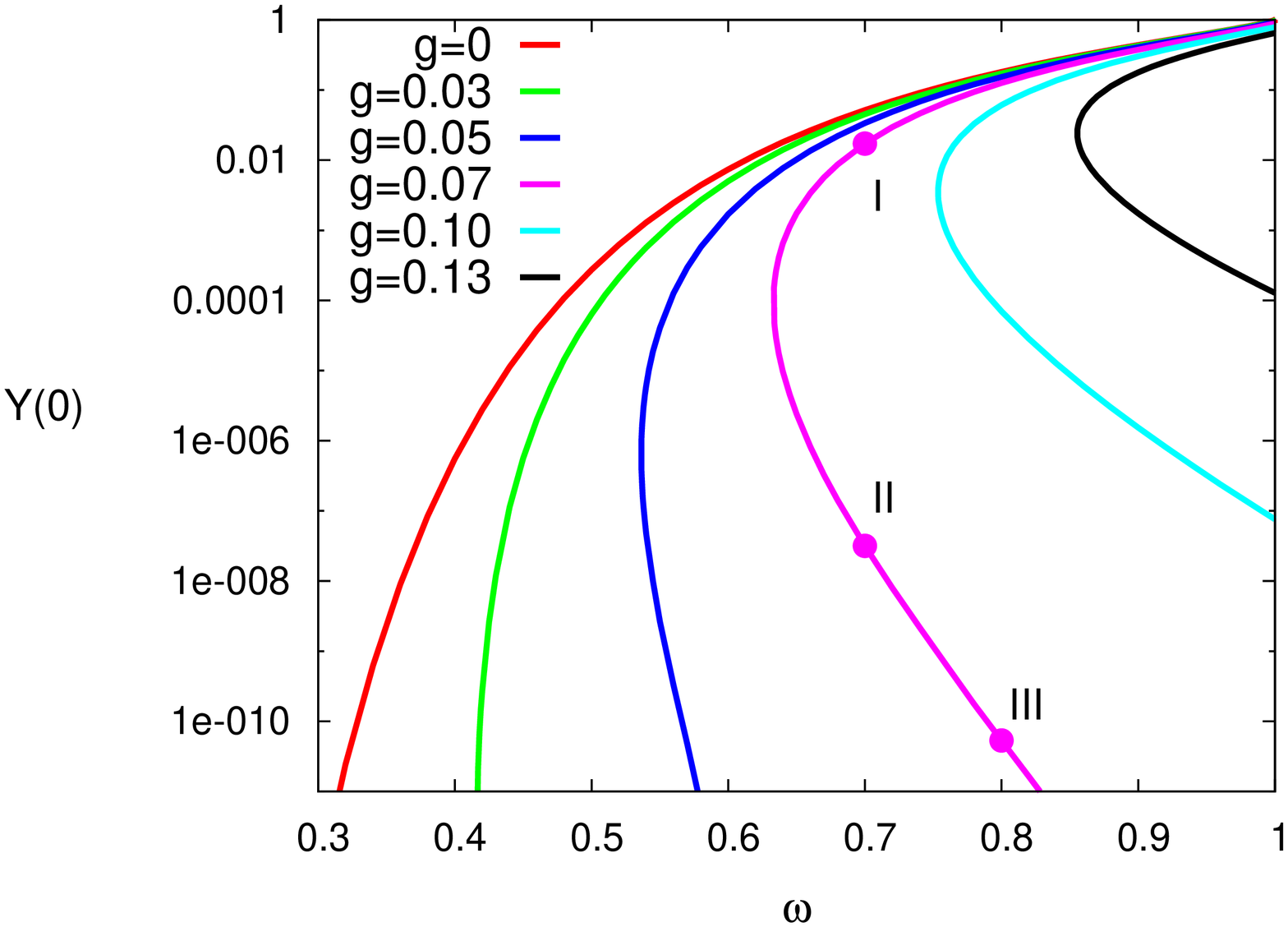}
\end{center}
\caption{\small $U(1)$ gauged Q-balls in the model \re{lag-fls}:
The total energy of the solutions (upper left plot), the values of
the function $C(0)$ (upper right plot), the gauge potential
$A_0(0)$ (bottom left plot), the scalar profile functions $X(0)$
(bottom middle plot) and $Y(0)$ (bottom right plot) at $r=0$ are displayed as functions of the
angular frequency $\omega$ for some set of values of the gauge
coupling $g$.}
    \lbfig{fig5}
\end{figure}

%

\begin{figure}[h!]
\begin{center}
\includegraphics[height=.32\textheight,  angle =-90]{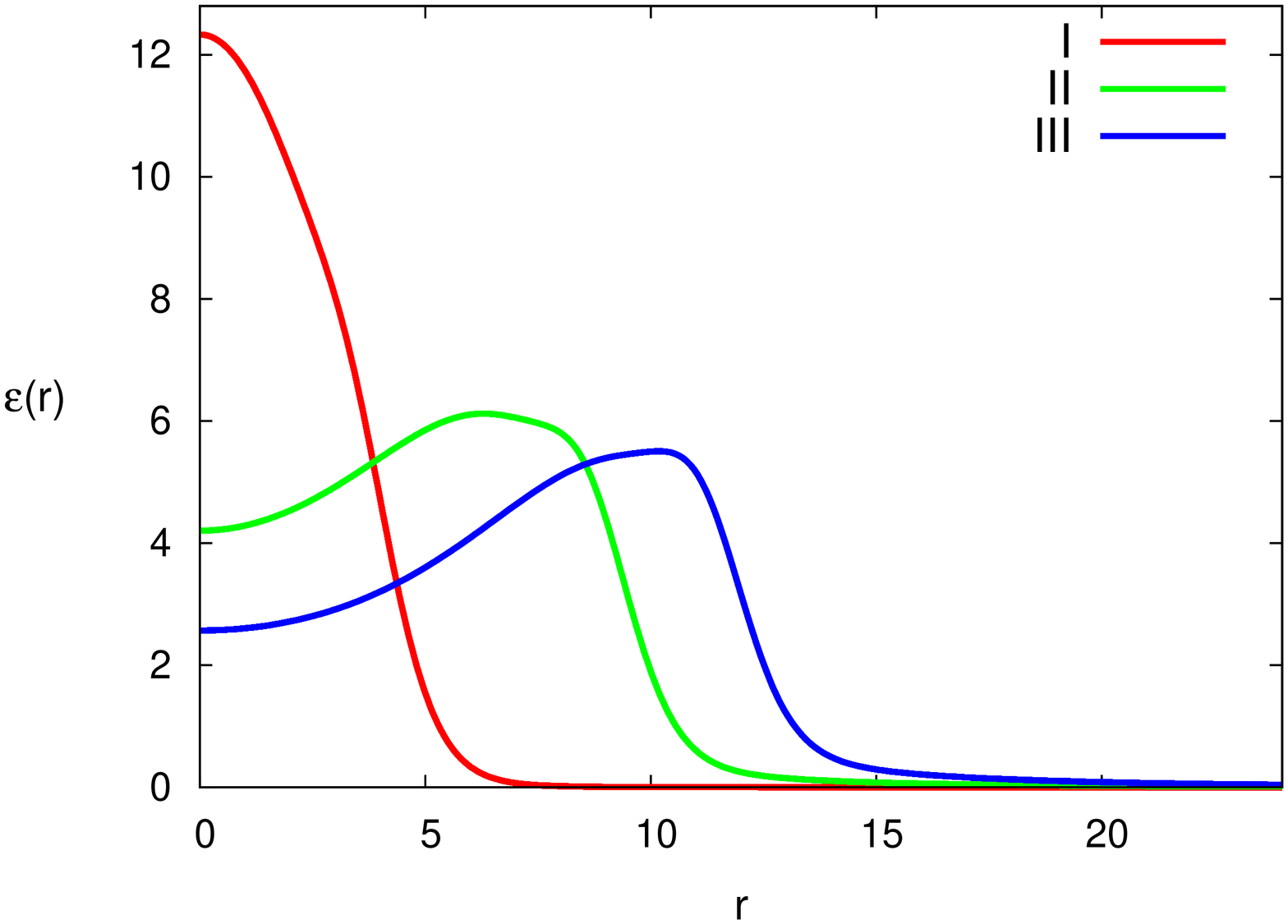}
\includegraphics[height=.32\textheight,  angle =-90]{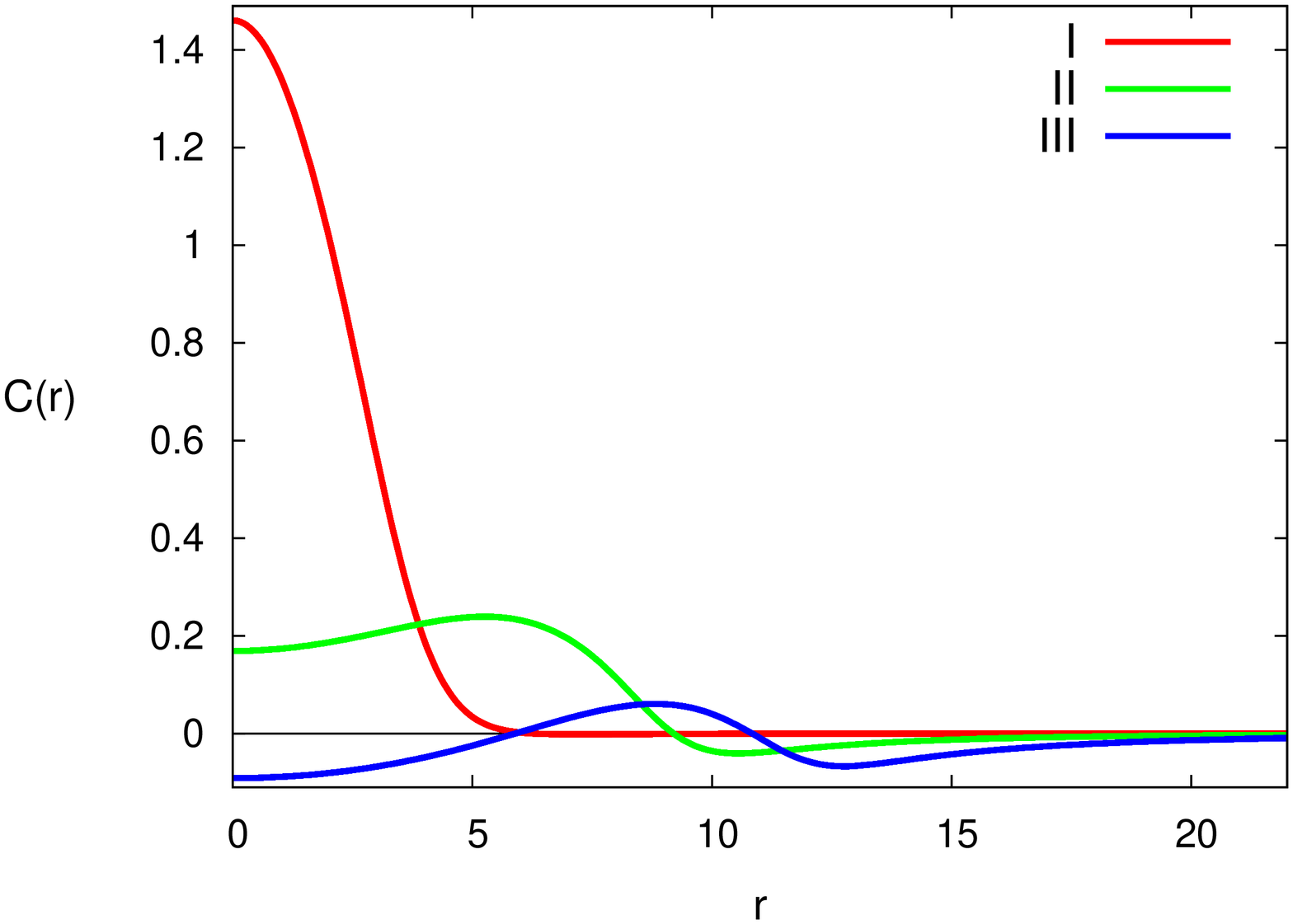}
\includegraphics[height=.32\textheight,  angle =-90]{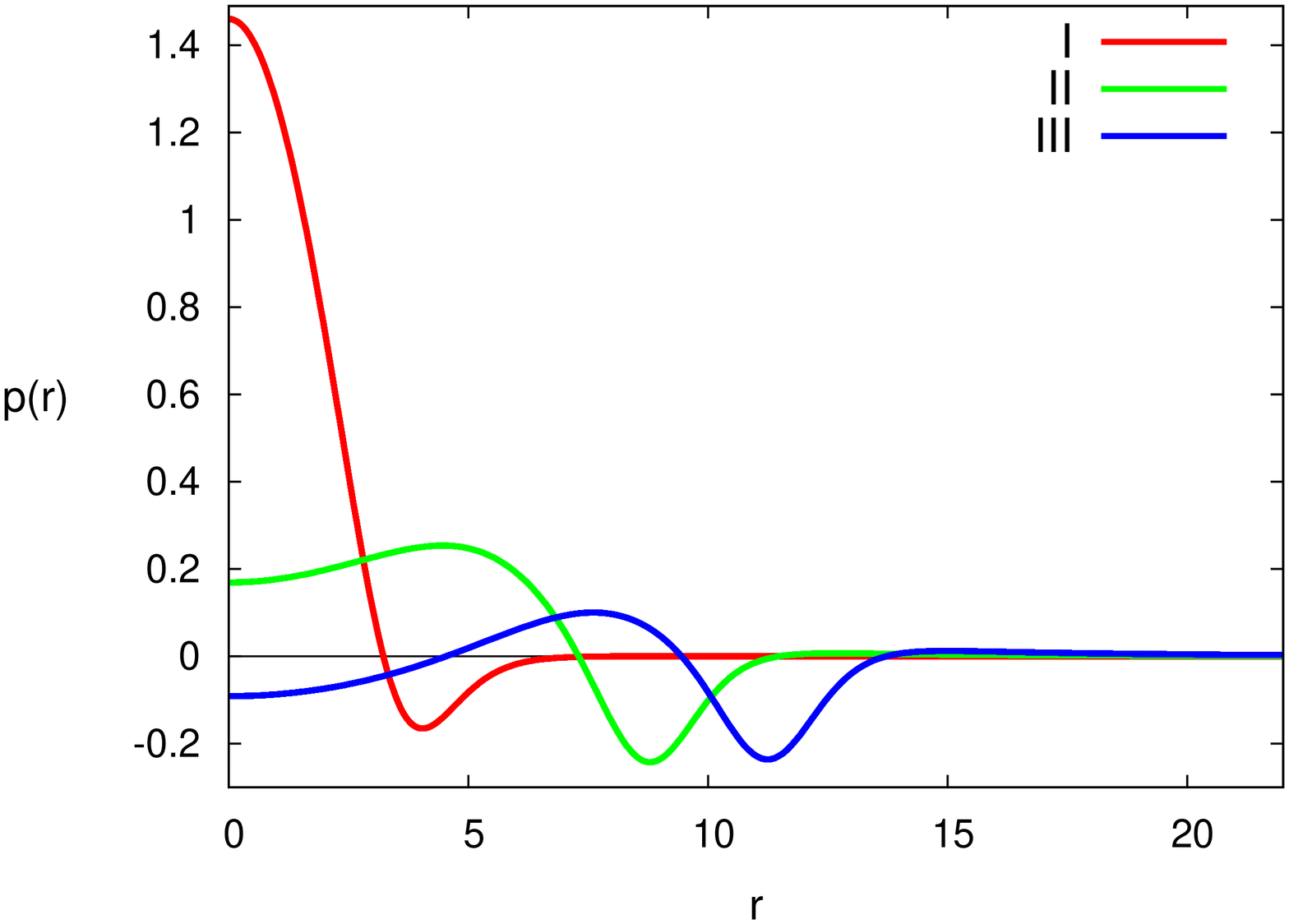}
\includegraphics[height=.32\textheight,  angle =-90]{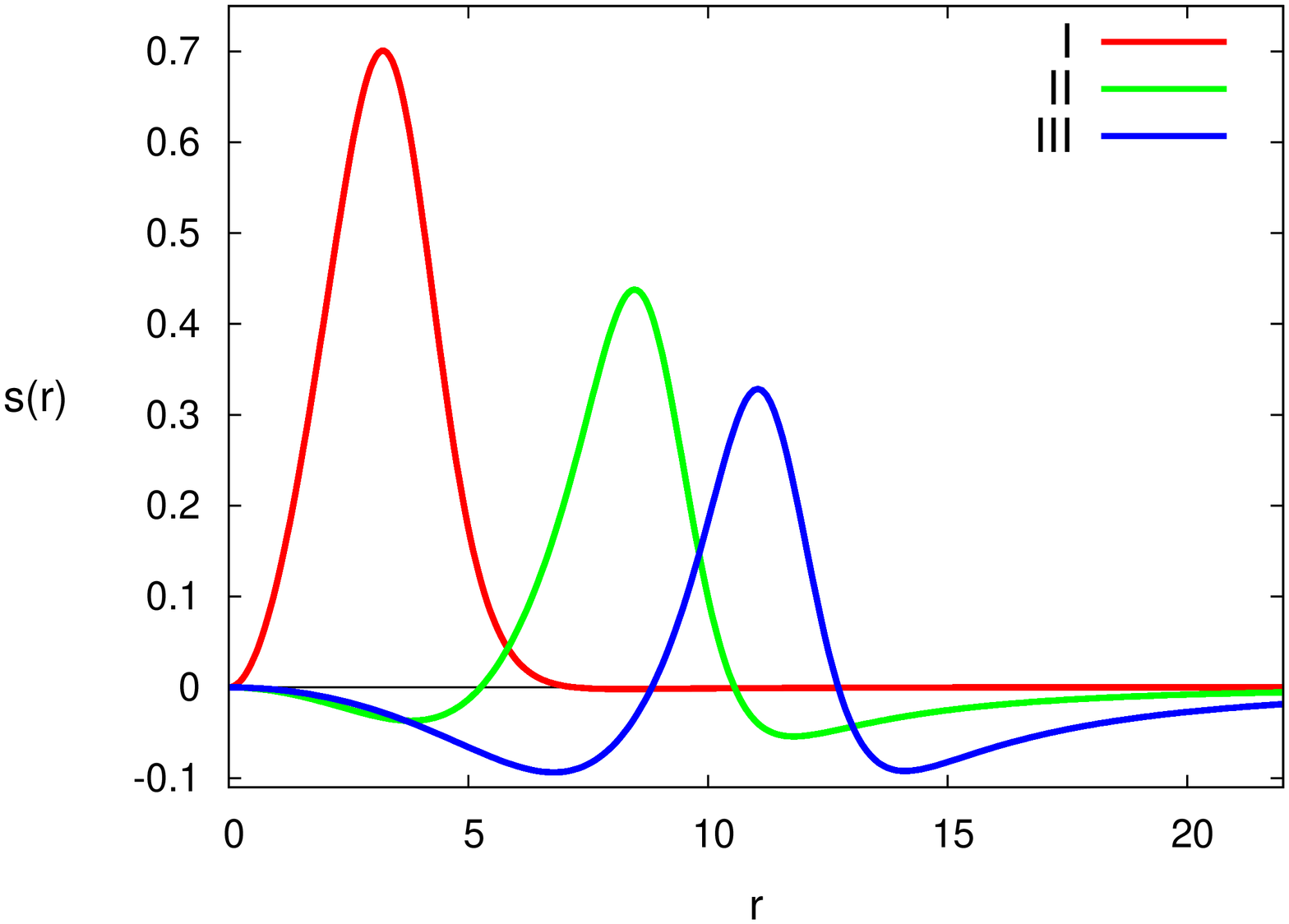}
\end{center}
\caption{\small
Solutions of the model \re{lag-fls}, labelled as points $I,II$ and $III$
on the Fig.\ref{fig5}: The distributions of the  energy density $\varepsilon(r)$
(upper left plot), the function $C(r)$ (upper right plot), the pressure function  $p(r)$
(bottom left plot) and the shear force function $s(r)$ (bottom right plot).}
    \lbfig{fig8}
\end{figure}

Spherically symmetric solutions of the model \re{lag-fls}
have been studied before \cite{Lee:1991bn}. The general
pattern is that, by analogy with the one component model \re{lag-Coleman}, the $U(1) $gauged
Q-balls exist for a restricted domain of values of the
parameters of the system.  The repulsive electromagnetic interaction reduces the allowed range
of values of the angular frequency $\omega$.
Note that, in the decoupled limit $g=0$, the ordinary
Friedberg-Lee-Sirlin Q-balls exist for all non-zero values of  scaled frequency
$\omega \in [0,\omega_\mathrm{max}]$. where the upper critical value corresponds to the mass of
free charged quanta of scalar excitations, $\omega_\mathrm{max}=m$. In our numerical simulations we set $m=1$.

For $\mu \neq 0$, there are
two branches of $E(\omega)$ curves with a bifurcation at $\omega= \omega_{cr}$
(see Fig.~\ref{fig5}, left upper plot). By analogy with the corresponding dependencies in the model
\re{lag-Coleman},
the energy of scalar interactions is dominating along the lower branch while the
electrostatic energy becomes much larger on the second forward branch (cf Fig.~\ref{fig1}).

It was noticed \cite{Kunz:2021mbm} that, as the angular frequency approaches the minimal critical value, the
real scalar component becomes very close to zero inside some domain
at the center of the Q-ball.  Within this  region both the complex scalar component $\phi$
and the gauge field $A_0$ are  massless. As the angular frequency increases along the second branch,
the dominating electrostatic interaction forms a
compact domain with a wall which separates the vacuum $\psi=1$  on the exterior and confining the massless
fields in the interior. The gauged Q-ball rapidly inflates along the second branch,
however both the total energy and the charge remain finite as $\omega \to \omega_\mathrm{max}$.

This pattern is not much different from that discussed above for the gauged Q-ball
in the one-component model \re{lag-Coleman}, although the range of allowed values of the gauge
coupling constant $g$ can be different. Similarly, the stability criteria \re{criterion}
becomes violated on the second branch, see Fig.\ref{fig8}.

The situation changes dramatically in the massless limit $\mu =0 $ \cite{Levin:2010gp,Loiko:2018mhb}.
This is a case of "hairy" gauged Q-balls with long-range real scalar component.
In such a case the second (upper)
branch disappears and both the total energy and the charge of the configuration
increase monotonically as $\omega$ decreases. Notably, they both tend to zero for $\omega \to \omega_\mathrm{max}=m$ and
diverge at some critical minimal value of the angular frequency $\omega_\mathrm{cr}$ \cite{Loiko:2019gwk}, see
Fig.~\ref{fig6}, upper left plot. This value increases with the gauge coupling, in the decoupled limit $g=0$ the
ungauged massless Friedberg-Lee-Sirlin Q-balls exist for the whole range of values of the angular frequency $\omega \in [0,1]$.
Remarkably, the "hairy" Q-balls are always stable with respect to linear perturbations \cite{Loiko:2018mhb}.

\begin{figure}[h!]
\begin{center}
\includegraphics[height=.33\textheight,  angle =-90]{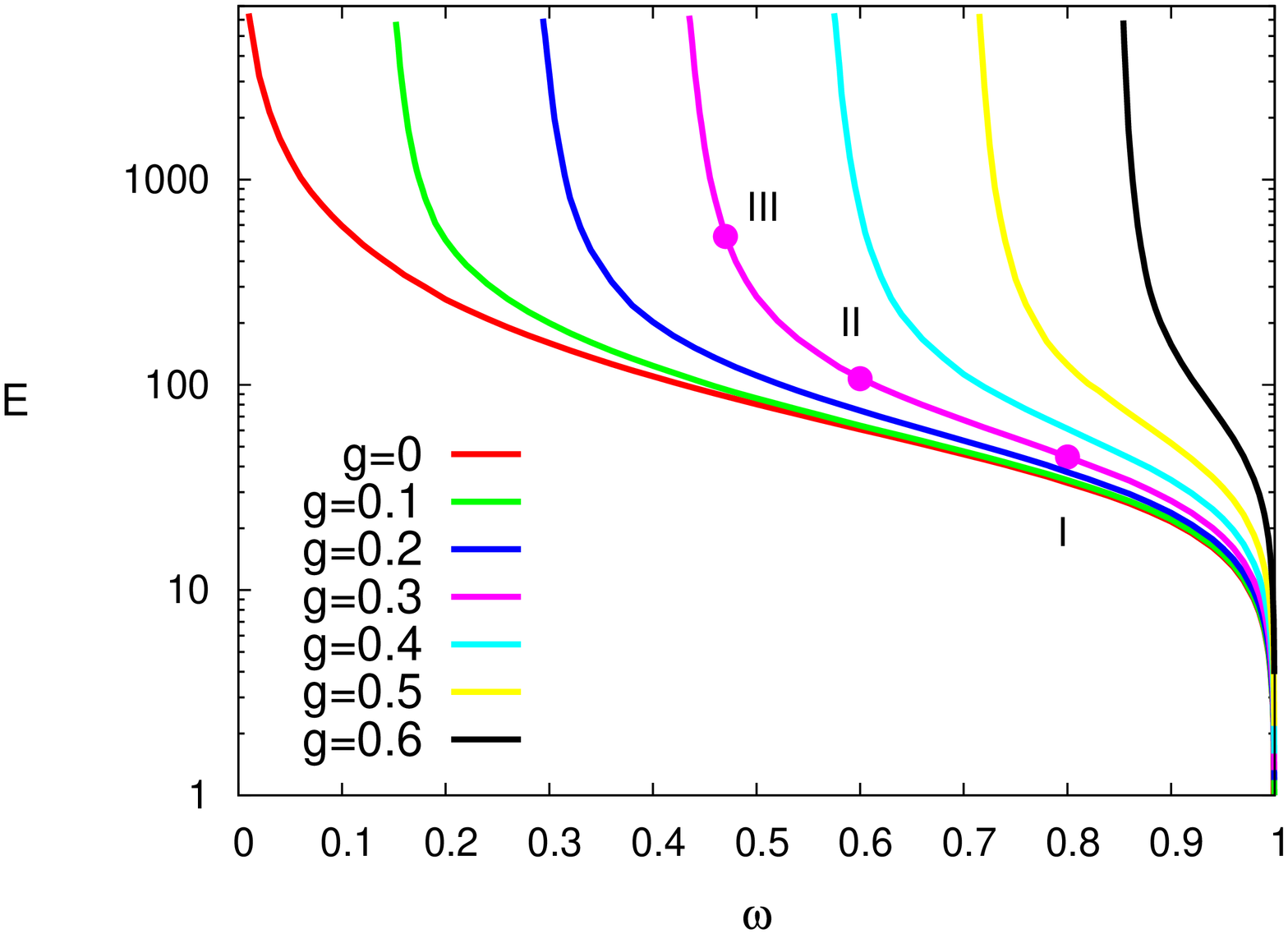}
\includegraphics[height=.33\textheight,  angle =-90]{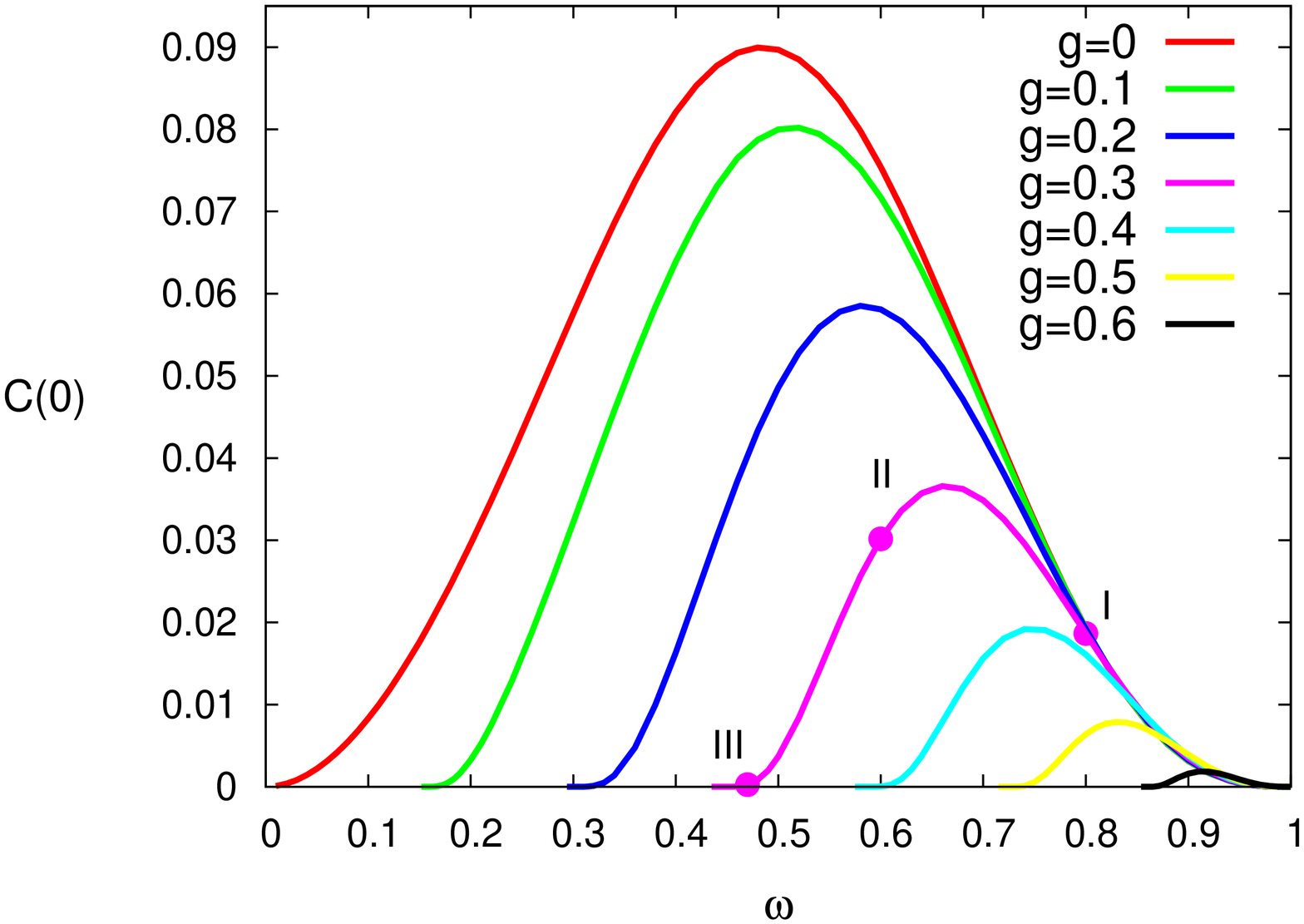}
\includegraphics[height=.24\textheight,  angle =-90]{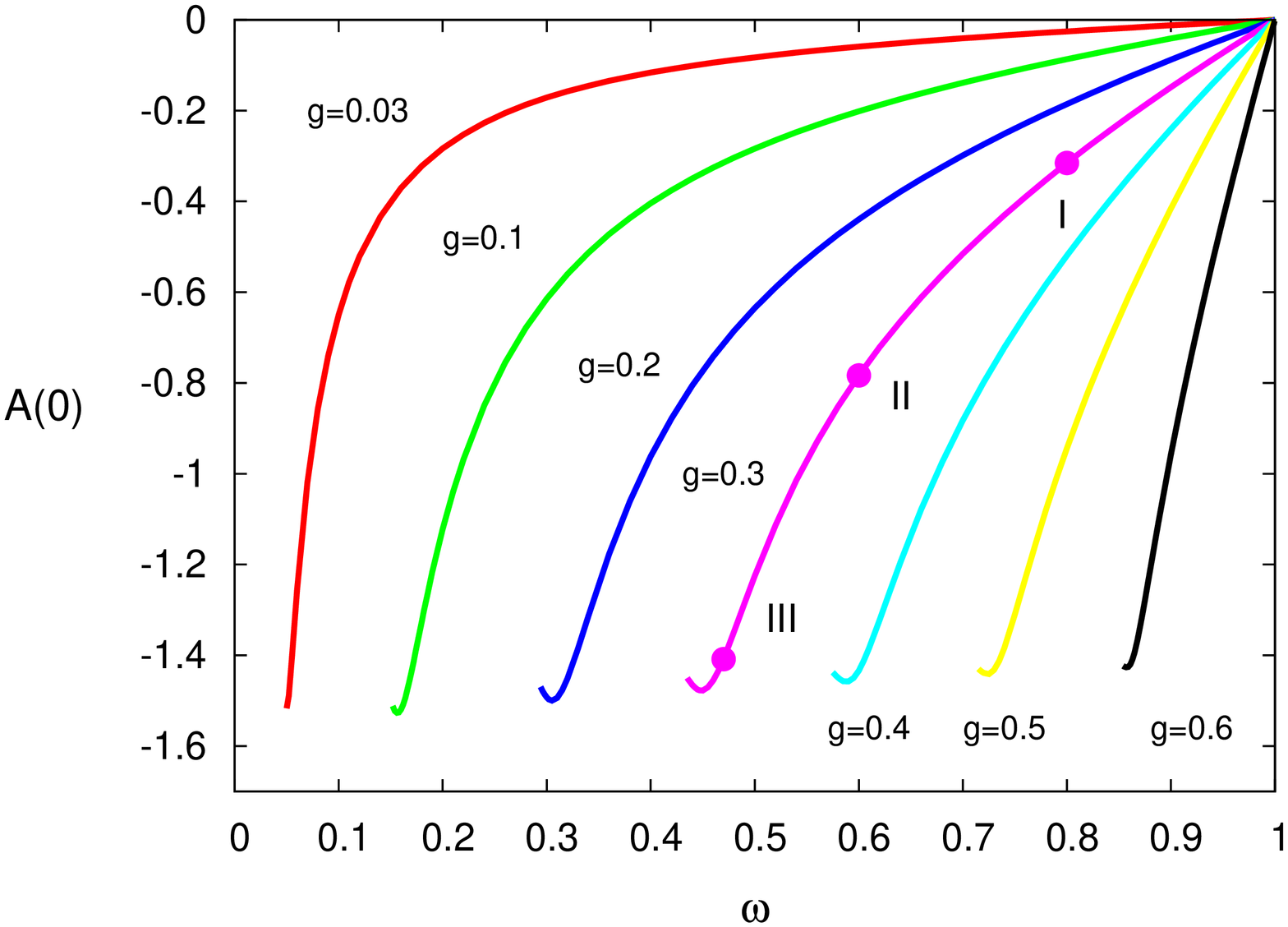}
\includegraphics[height=.24\textheight,  angle =-90]{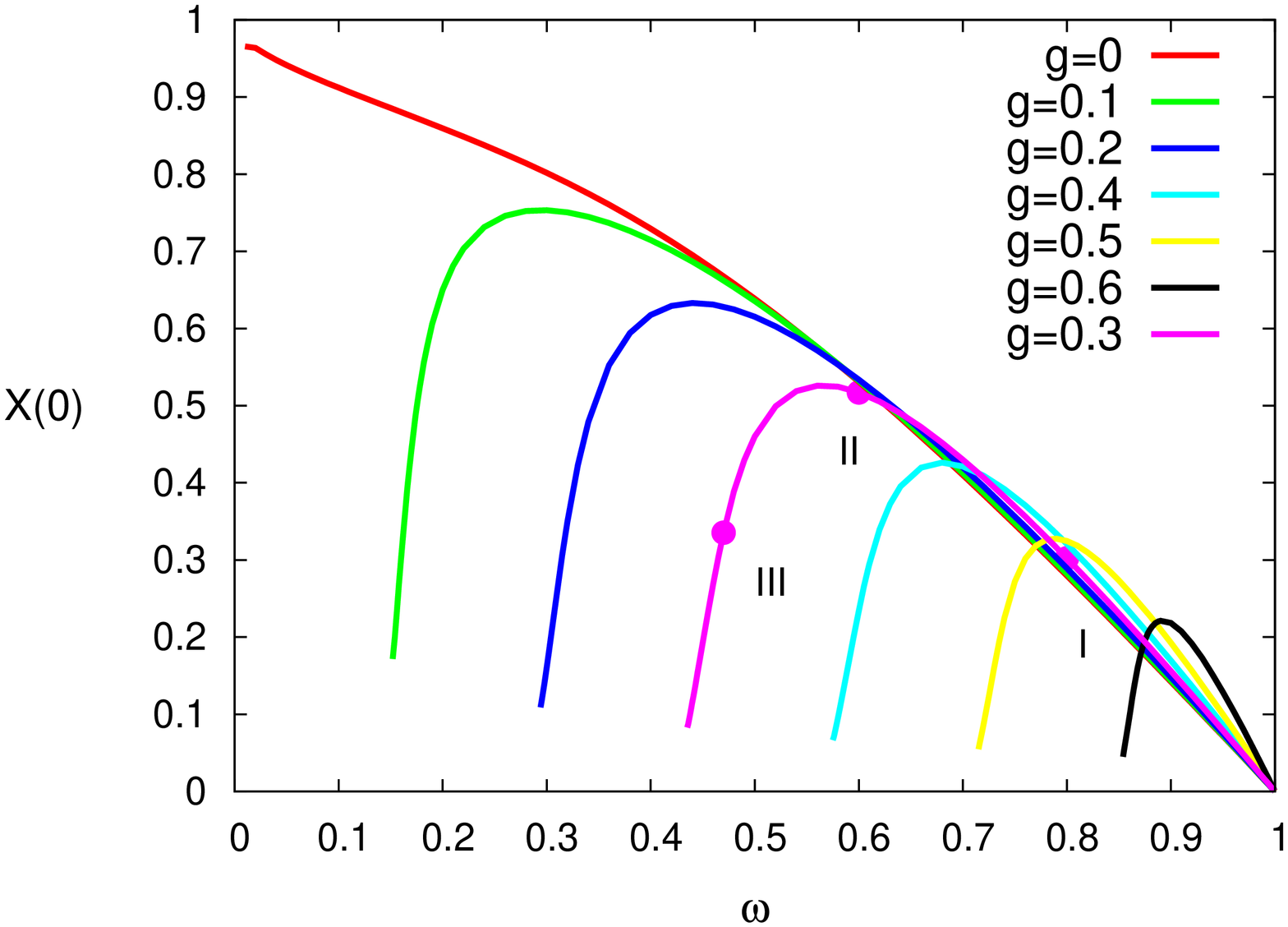}
\includegraphics[height=.24\textheight,  angle =-90]{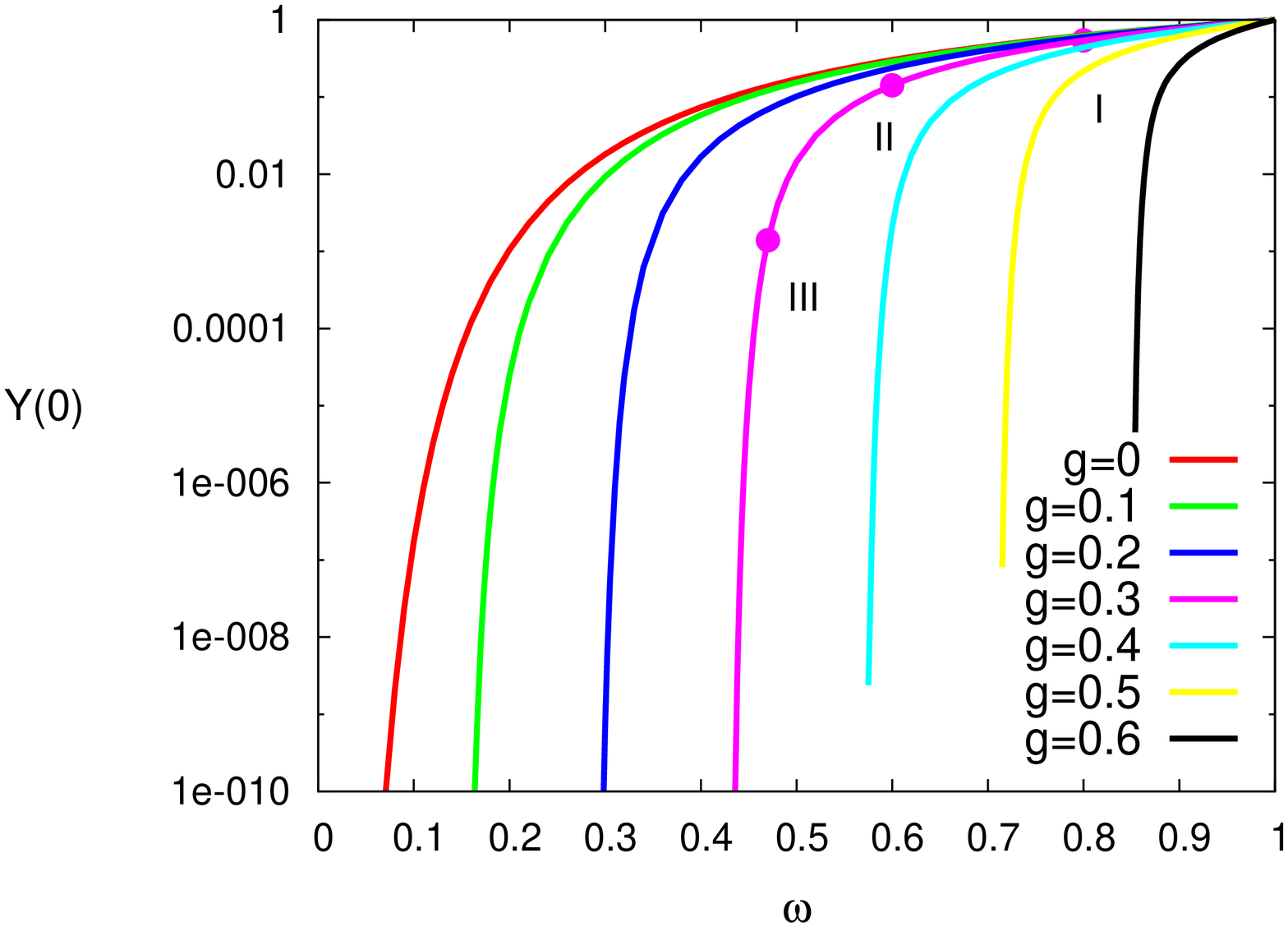}
\end{center}
\caption{\small
$U(1)$ gauged "hairy" ($\mu=0$) Q-balls in the model \re{lag-fls} : The total energy of the solutions (upper left plot), the values of the function $C(0)$ (upper right plot), the gauge
potential $A_0(0)$ (bottom left), and the scalar profile functions $X(0), Y(0)$ at $r=0$  (bottom middle and right plots, respectively) are displayed as
functions of the angular frequency $\omega$ for some set of values of the gauge coupling $g$.}
    \lbfig{fig6}
\end{figure}
\begin{figure}[h!]
\begin{center}
\includegraphics[height=.32\textheight,  angle =-90]{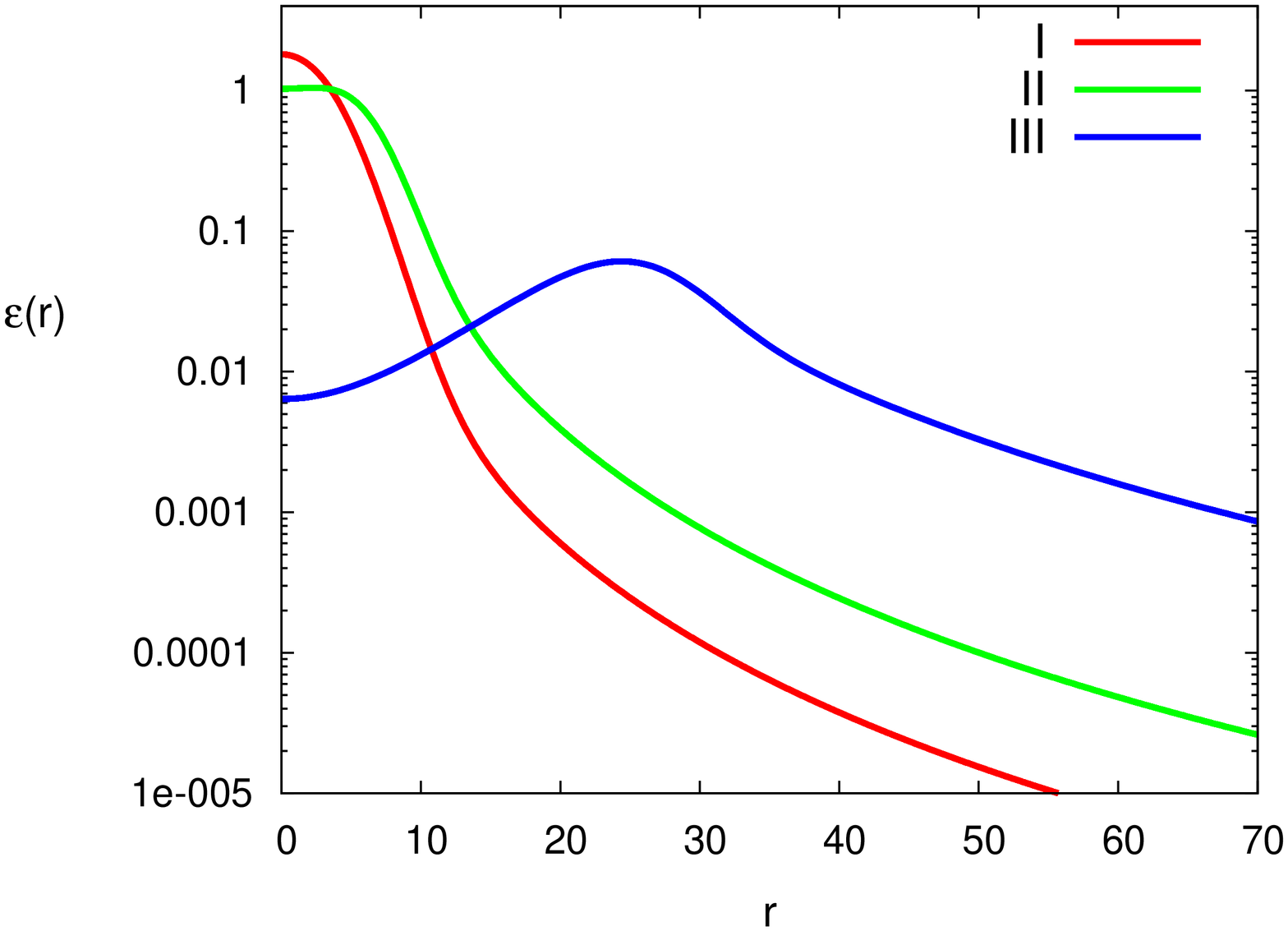}
\includegraphics[height=.32\textheight,  angle =-90]{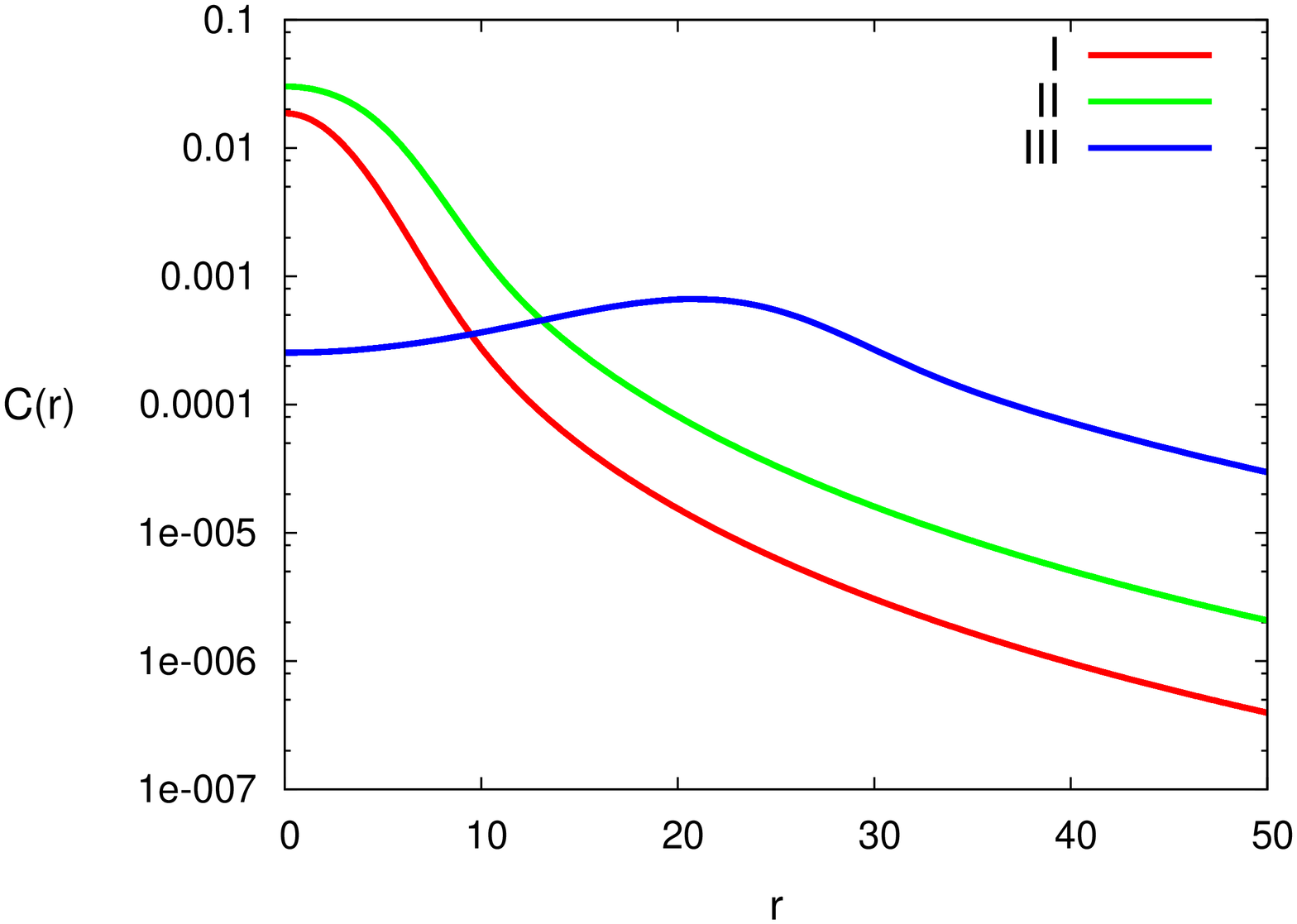}
\includegraphics[height=.32\textheight,  angle =-90]{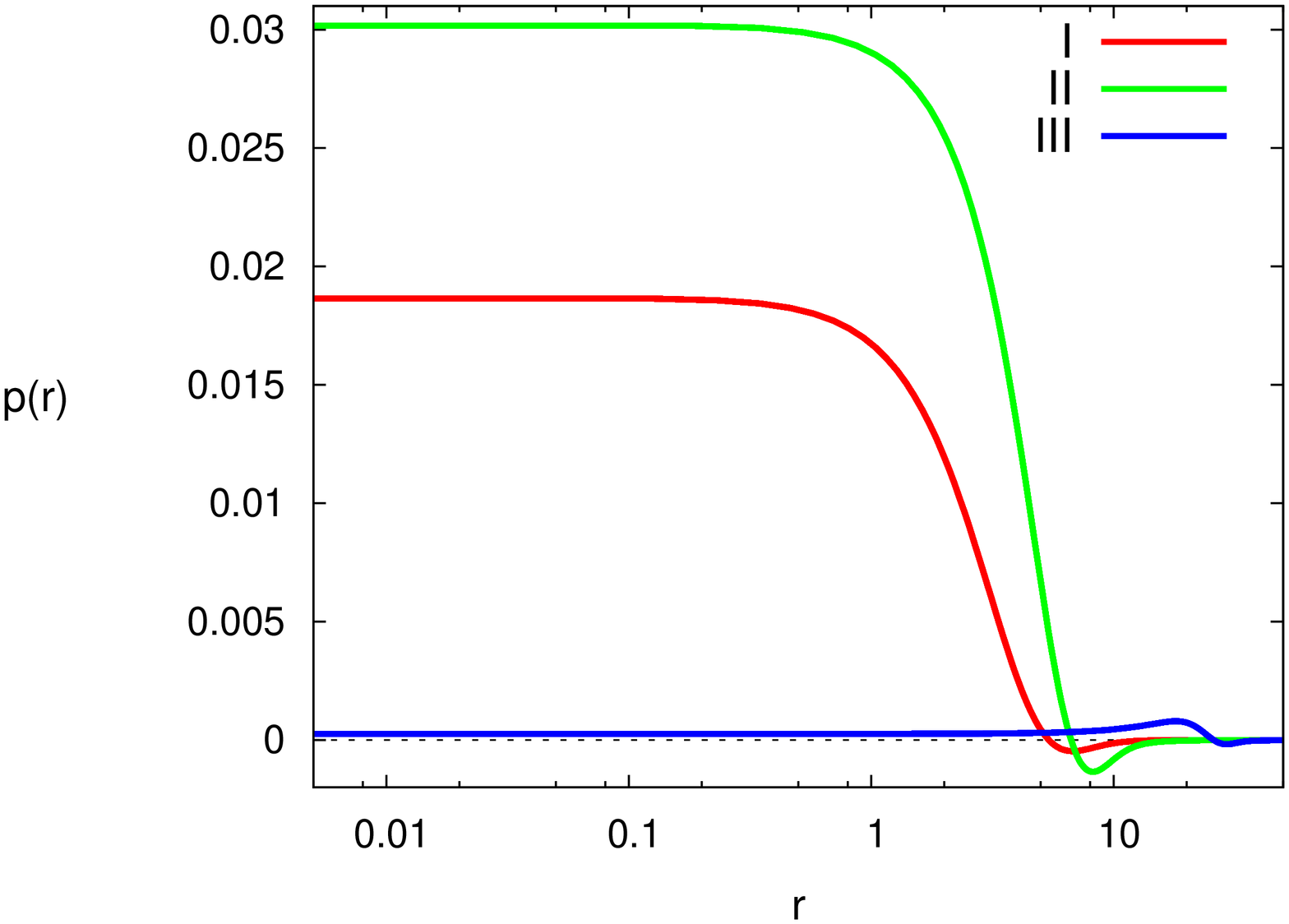}
\includegraphics[height=.32\textheight,  angle =-90]{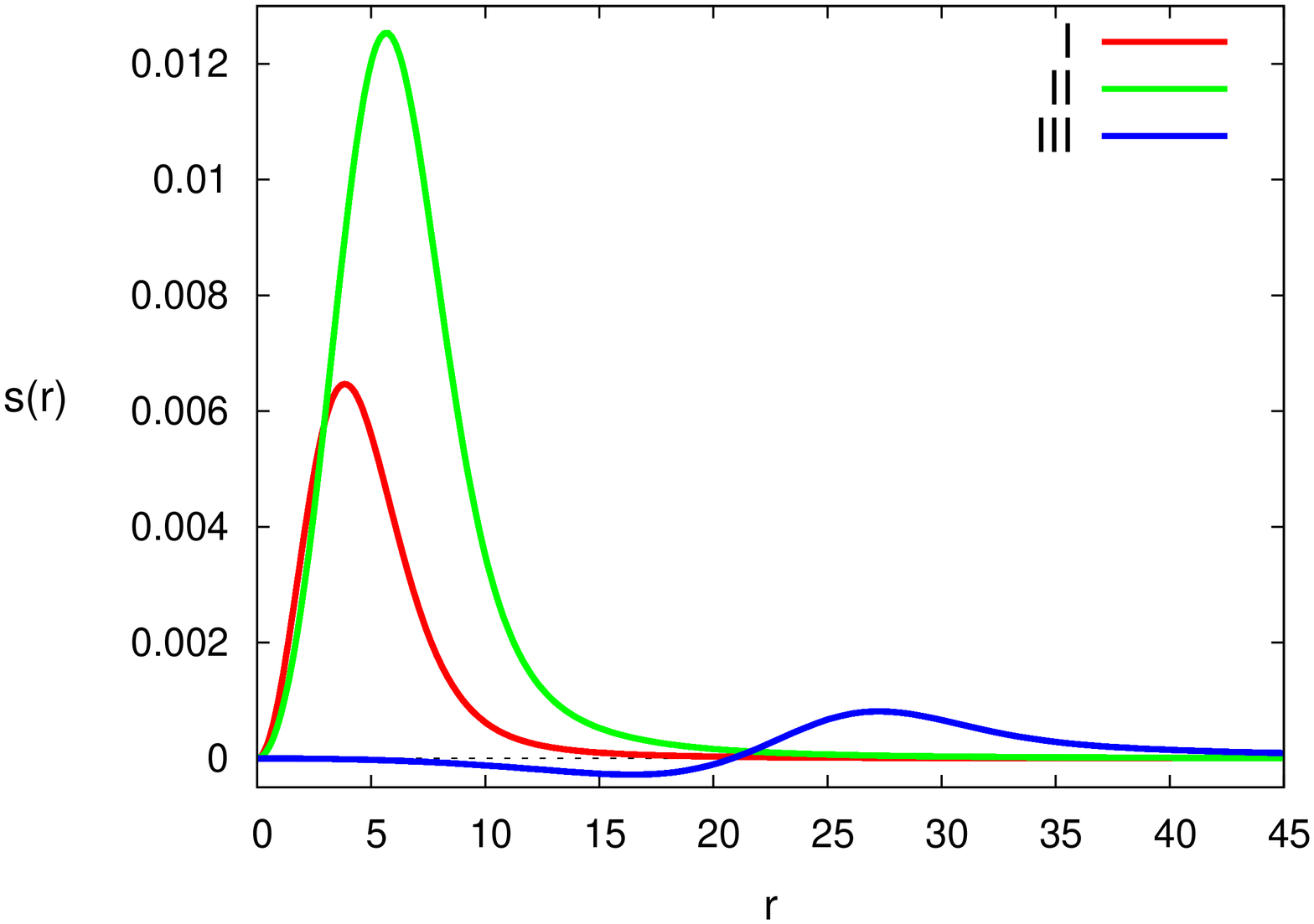}
\end{center}
\caption{\small
Solutions of the model \re{lag-fls}, labelled as $I,II$ and $III$ on the Fig.\ref{fig6}: The distributions of the  energy density $\varepsilon(r)$  (upper left plot), the function $C(r)$ (upper right plot), the pressure function  $p(r)$ (bottom left plot) and the shear force function $s(r)$ (bottom right plot).}
    \lbfig{fig7}
\end{figure}

Indeed, numerical results confirm that the function $C(r)$ \re{cr-fls} always remain positive as $\mu=0$, see Fig.\ref{fig7}, upper left plot. In this figure we also displayed the corresponding profiles of the energy density distribution $\epsilon (r)$ \re{eng-fls}, the pressure function $p(r)$ and the shear force $s(r)$, defined as \re{p-s-fls}, of the particular "massless" solutions, labelled as $I,II$ and $III$ on the Fig.\ref{fig6}, respectively.


\section{Conclusions}
In the present paper, we revisited
the problem of classical stability of
$U(1)$ gauged Q-balls in a non-renormalizable one-component scalar model with a sixtic potential and in the
two-component Maxwell-Friedberg-Lee-Sirlin model with symmetry breaking potential. Our approach is based on a treatment of the
interior of a gauged Q-ball as an elastic medium and consideration of the corresponding
matrix elements of energy momentum tensor, which contains information about spatial distribution of internal
forces \cite{Mai:2012yc,Polyakov:2002yz,Perevalova:2016dln,Polyakov:2018zvc}. This  analysis supplements the well-known
classical Vakhitov-Kolokolov stability criterion \cite{Vakhitov}, previously modified for the $U(1)$ gauged Q-balls \cite{Panin:2016ooo}.
We derived the expressions for the distributions of the  pressure and shear forces, acting in the interior of the
gauged Q-balls and showed that the von Laue stability condition is always satisfied. Further, we analyse a local stability criteria,
suggested previously in studies of hadrons. We show that this criteria becomes violated on the second forward branch of solutions,
as the electrostatic energy becomes much larger than the energy of scalar interactions.
We would like to emphasize that the stability condition \re{criterion}
is stronger than the usual relation between the mass of a Q-ball and the mass of $Q$ free scalar exitations, as it becomes violated on the
upper branch, the Q-ball may not decay into radiation but rather may emit some part of energy in a transfer to the lower branch.

On the other hand, the
inequality \re{criterion} is always satisfied for ungauged Q-balls \cite{Mai:2012yc,Mai:2012cx}. Similarly, the results of
our numerical simulations demonstrate that the $U(1)$ gauged Q-balls in the two-component Friedberg-Lee-Sirlin-Maxwell model
are stable for all range of values of the parameters of the system in the massless limit as there is just one branch of solutions.

It should be noted that, unlike the Vakhitov-Kolokolov stability criterion \cite{Vakhitov}, the inequality \re{criterion} is not
related with a perturbative consideration of spectra of linearized perturbations of a soliton. This inequality follows from
an a naive approximation of a Q-balls as a continuous media and  related concepts of pressure and shear forces,
which is not always well justified. An interesting question is if the inequality
\re{criterion} can be obtained as an effective relation derived from the quantum microscopic theory.
It would be interesting to clarify, if the suggestion about possible relation of the criterion \re{criterion}
and the restriction on the speed of sound \cite{Polyakov:2018zvc,Polyakov:2018rew} is correct. Another interesting task is to study
full 3+1 dynamical evolution of the gauged Q-balls for all range of values of the parameters. We hope
to address these questions  in our future work.

This work is dedicated to the memory of Maxim Polyakov, a brilliant physicist and a very close friend of one of us (YS).
It originates from discussions with Maxim without whom this article would never have been written.

 \begin{small}
 
 \end{small}

 \end{document}